\documentclass[pra,twocolumn,amssymb,amsmath]{revtex4-1}
\usepackage{graphicx}
\usepackage{epsfig,color}
\usepackage{amsmath,amssymb}
\begin{document}

\title{  Quantum simulations of gauge theories with ultracold atoms:  \\ local  gauge invariance from angular momentum conservation }
\date{\today}

\author{Erez Zohar$^1$, J. Ignacio Cirac$^2$, Benni Reznik$^1$}

\begin{abstract}
\begin{center}
$^1$School of Physics and Astronomy, Raymond and Beverly Sackler
Faculty of Exact Sciences, Tel Aviv University, Tel-Aviv 69978, Israel.

$^2$Max-Planck-Institut f\"ur Quantenoptik, Hans-Kopfermann-Stra\ss e 1, 85748 Garching, Germany.
\end{center}
Quantum simulations of High Energy Physics, and especially of gauge theories, is an emerging and exciting direction in quantum simulations. However, simulations of such theories, compared to simulations of condensed matter physics,
must satisfy extra restrictions, such as local gauge invariance and relativistic structure. In this paper we discuss these special requirements, and present a new method for quantum simulation of lattice gauge theories using
ultracold atoms. This method allows to include local gauge invariance as a \emph{fundamental} symmetry of the atomic Hamiltonian, arising from natural atomic interactions and conservation laws (and not as a property of a low
energy sector). This allows us to implement elementary gauge invariant interactions for three lattice gauge theories: $U(1)$ (compact QED), $\mathbb{Z}_N$ and $SU(N)$ (Yang-Mills), which can be used to build quantum simulators in
$1+1$ dimensions.
We also present a new "loop method", which uses the elementary interactions as building blocks in the effective construction of quantum simulations for $d+1$ dimensional lattice gauge theories ($d>1$), but unlike in
previous proposals, here gauge invariance and Gauss's law are natural symmetries which do not have to be imposed as a constraint. We discuss in detail the quantum simulation of $2+1$ dimensional compact QED and provide
a numerical proof of principle. The simplicity of the already gauge invariant elementary interactions of this model suggests it may be useful for future experimental realizations.
\end{abstract}

\maketitle

\tableofcontents

\section{Introduction}

The concept of quantum simulations goes back to the 1980s, when Richard Feynman suggested
the possibility to simulate Quantum Mechanics using quantum computers \cite{Feynman1982}. Over the recent
decades, the extraordinary progress both in theory and experiments, has enabled an
unprecedent control in systems like cold atoms \cite{Lewenstein2012,Bloch2012}, or trapped ions \cite{Blatt2012}. This control allows
us nowadays to use them to simulate other quantum systems in an analog way, in the spirit of
Feynman's visionary ideas. This has opened up a path to observe and understand many physical
phenomena which are, at least currently, unreachable in the context of analytic calculations or experimental
measurements of the original systems \cite{CiracZoller}.

Of specific interest are quantum simulations of many-body quantum models that appear in condensed
matter physics. Some of those models are difficult to handle even with the most advanced numerical
techniques, and thus quantum simulation appears as an important tool to investigate them.
These models include, for example, Hubbard models, Heisenberg-like models in different lattice
geometries. Atomic systems provide us with a natural playground to simulate those models,
since once they are loaded in optical lattices, they are basically described in terms of simple
Hubbard models whose properties can be tuned with external fields. During the last years,
many quantum simulations of such models have been proposed, different techniques have been
developed, and some of them have even been realized experimentally.

Gauge theories constitute the fundamental building blocks of the standard model of High Energy
Physics (HEP). They are built up out of fermionic and bosonic particles (or fields), which represent
matter and the force carriers, respectively. As a many-body quantum system, they are extremely
rich in intriguing phenomena, and in some limits are very hard to study, even with the most
advanced numerical techniques. Thus, a natural question is whether one could use the existing
quantum simulators based on atoms in order to observe such phenomena, as well as to investigate
regimes where standard techniques do not work. Unlike with condensed matter problems, however,
the simulation of HEP models does not appear in a natural way in such atomic systems. In particular,
they require: (a) Inclusion of both fermions and bosons; (b) interactions preserving local gauge invariance,
which results in Gauss's law, as well as (c) Lorentz invariance (at least in the proper
continuum limit).

In recent years, several works have proposed to use cold atoms in optical lattices to simulate
HEP theories \cite{Cirac2010,Bermudez2010,Boada2011,Kapit2011,Zohar2011,Mazza2012,Boada2012,Zohar2012,Banerjee2012,Zohar2013,Tagliacozzo2013,NA,Rishon2012,TagliacozzoNA}. Most of them propose to use optical lattices to
enforce (c), and perturbation theory such that the desired interacting
terms appear in the low energy sector. In some cases, this may lead to undesirable effects, since
the gauge invariance is not exact, or since one has to go to high order perturbation theory, which
leads to very weak effective interactions and strong constraints.

In this paper, building on our early work on quantum simulation of HEP models \cite{Zohar2011,Zohar2012,Topological,Zohar2013,NA}, we introduce
new techniques in order to implement the conditions (a-c) above in an optimal manner. First,
we propose to use several internal states of the fermionic and bosonic atoms such that
the gauge invariance of the resulting HEP model is a direct consequence of the conservation
of angular momentum in the original atomic scattering processes, and not a property of the low
energy sector after perturbation theory. Second, since  the original Hamiltonian is already
gauge invariant,  Gauss's law does not have to be enforced: given that (in operator form) it
commutes with the Hamiltonian, one just has to initialize the atoms in a state which satisfies
it and then the dynamics will always occur in the subspace fulfilling that law. Thus, one could start out
with the parameters corresponding to a regime where the ground state is well defined, and then
turn adiabatically the physical parameters in order to explore other regimes. Third, we propose to
engineer the lattice system such that the traps for the bosonic atoms lie between the traps for the
fermionic ones. In this way, fermionic hopping is mediated by a collision with the bosonic atoms, giving
rise to the matter-gauge field interactions with the largest possible coefficient, since the overlap integral
responsible for this term is maximal. Fourth, we provide a new method (the "loop method") to construct the plaquette
interactions that give rise to the dynamics of the gauge field (in the case of 2+1 dimensions). As in
previous suggestions, we use perturbation theory in order to obtain the effective terms in fourth order. However, we
show that lower order terms in the perturbation series only renormalize our theory, so that we obtain
the desired plaquette terms under the conditions that are equivalent to a second order perturbation
(and not to fourth order). Furthermore, unlike in previous proposals, they are
constructed out of already gauge invariant objects, and thus they do not require the explicit use of the
Gauss law.

The paper is organized as follows: in section II we put down the basics required for such quantum simulations: we briefly discuss gauge theories and lattice gauge theories, and deduce the requirements that quantum simulations
of such systems must fulfill; We discuss simpler high energy physics models which already show the interesting physics, but are simpler for quantum simulation, and review previous suggestions for quantum simulations of such systems.
In section III we describe the simulating system - the general structure of the optical lattice and atomic Hamiltonian needed for such simulations, and then, in section IV, we show how to get, in the fundamental Hamiltonian (without perturbation theory) the gauge invariant elementary interactions on links, for several gauge theories $U(1)$ , $\mathbb{Z}_N$
 and $SU(N)$. Then, in section V, we utilize these elementary interactions to build quantum simulations of $1+1$ dimensional models -
$U(1)$ (a full nonperturbative simulation of the $1+1$ dimensional Schwinger model) and $SU(N)$. In section VI we introduce the new "loop method" and show how to use the gauge invariant nonperturbative elementary interactions to construct effectively plaquette interactions, required for the simulation of models in more than $1+1$ dimensions, for these
three gauge theories, and in section VII we show how to use them and construct quantum simulators for $2+1$-d lattice gauge theories ($U(1)$, $\mathbb{Z}_N$
and $SU(N)$. The paper also contains an appendix, briefly expanding on some properties of the gauge theories whose quantum simulations are discussed.

\section{Quantum Simulation of High Energy Physics}

\subsection{Basics of High Energy Physics}
\label{bHEP}
The standard model of High Energy Physics (HEP) is a Quantum Field Theory (QFT), in which the elementary particles can be divided into two separate groups. Matter particles (quarks and leptons) are fermions, represented by Dirac
fields, while the force mediators, which are responsible to the interactions among matter particles, are gauge bosons.
Being a \emph{gauge} boson, the gauge field must satisfy a special symmetry, which is \emph{Local Gauge Invariance}. This symmetry may be either manifest or broken, but it is the nature of this symmetry which is responsible to the
very special coupling between matter and gauge fields. Each gauge theory is based on a gauge group, whose elements are fundamental objects of the theory, forming the \emph{group space} in which the gauge transformations apply: these
transformations do not correspond to changes in any physical observable, and thus they form a symmetry. The gauge groups may be either continuous or discrete, abelian and non-abelian.

Both in abelian and non-abelian theories the gauge fields are massless \footnote{In this work we disregard the Higgs mechanism, which breaks the gauge symmetry and gives masses to the gauge fields, and also introduces a \emph{scalar} (bosonic) matter field.}. However, in abelian theories the gauge fields are chargeless, while in non-abelian theories they carry charge. Quantum Electrodynamics (QED), for example, is an abelian ($U(1)$) gauge theory, whose charge is just the ordinary electric charge, and its gauge bosons, the photons, carry no charge. Quantum Chromodynamis (QCD), describing the strong interactions, is a non-abelian ($SU(3)$) gauge theory, whose charge is the \emph{color} charge, carried by the quarks but also by the gauge bosons - gluons - due to the non-abelian nature of the theory.
Abelian theories yield linear equations of motion (Maxwell's equations for QED), since the chargeless gauge bosons do not interact among themselves.
In non-abelian gauge theories, there are such self-interactions, due to the non-abelian charge carried by the gauge bosons, and this results in nonlinear equations of motion - the fundamental theory is described by the Yang-Mills
equations for an $SU(N)$ gauge theory \cite{Yang1954}. Furthermore, non-abelian theories are responsible for long-range forces, manifested, for example, by the electromagnetic Coulomb law. However, non-abelian theories manifest the
effect of \emph{confinement}, which binds matter particles together, such as quark confinement in $QCD$, which is responsible for the hadronic spectrum and forbids the existence of isolated free quarks \cite{Wilson,Polyakov}.
This non-perturbative phenomenon has been addressed over the recent decades in a variety of methods, including Lattice Gauge Theories (LGT), where the space-time is discretized, enabling a numerical Monte-Carlo
simulation \cite{Wilson,KogutSusskind,KogutLattice,Kogut1983}. However, such a classical simulation, although very useful for many things - finding the hadronic spectrum, for example - still faces problems such as the
sign-problem \cite{Troyer2005}, which makes it hard to approach the limit of many fermions (a finite chemical potential), and thus probing some exotical phases of gauge theories (for example, color superconductivity in $QCD$)
impossible using these methods \cite{Rischke2004,Alford2008,Fukushima2011}. Other than that, quantum simulations enables also the simulation of \emph{dynamics}, which is hard to simulate classically.

Another important feature of HEP is being a \emph{relativstic} theory, i.e. satisfying Lorentz invariance. This is of a great significance, as the interactions of elementary particles involve the regime of small distances and,
of course, high energies, which requires a relativistic treatment, which is mostly avoidable in the case of condensed matter physics. While this symmetry must be exactly met in the continuum limit, it cannot hold in a discretized
space-time as in LGTs. However, these theories still must include the proper remnants of Lorentz invariance, such that their continuum limit would be exactly relativistic.

\subsection{Basic ingredients of a Lattice Gauge Theory}
\label{BILGT}

\begin{figure}
  \centering
  \includegraphics[scale=1]{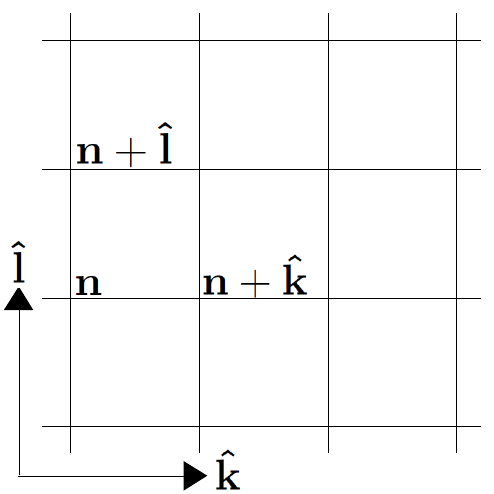}
  \caption{A part of the spatial lattice, in the $\mathbf{\hat k} - \mathbf{\hat l}$ plane. The labeling of the vertices is shown. The links are labeled by their source vertex and their direction. For example, the link connecting
  between the vertices $\mathbf{n}$ and $\mathbf{n + \hat k}$ is labeled as $\mathbf{n},k$. A spinor $\psi_{\mathbf{n}}$ is defined on each vertex $\mathbf{n}$, and a group element $U_{\mathbf{n},k}$ is defined on each link
  $\mathbf{n},k$.}
  \label{lattice}
\end{figure}

In lattice gauge theories, one can either discretize the entire (Euclidean) spacetime, or only the spatial directions. As we are interested in a Hamiltonian model, we shall use the latter latticization, introduced by Kogut and
Susskind \cite{KogutSusskind,KogutLattice,Kogut1983}.

In Hamiltonian LGTs, the fermionic (spinor) matter degrees of freedom, $\psi_{\mathbf{n}}$ reside on the vertices $\mathbf{n} \in \mathbb{Z}^d$ of a $d$-dimensional spatial lattice \footnote{Lattice fermions are a complicated
subject of its own, due to the problem of fermion doubling in the continuum limit. This problem can be resolved in several ways, but we shall not consider it here as it is irrelevant for our discussion.} (see figure \ref{lattice}).
These spinors may carry, generally, other indices, corresponding to possible physical quantum numbers of the matter fields, such as spin or flavor (which we avoid here) and also \emph{group space} indices, which we denote here by
lowercase roman letters, $a,b$, etc. And thus, generally
\begin{equation}
\psi_{\mathbf{n}} = \left( \psi_{\mathbf{n},a}\right) = \left(
\begin{matrix}
    \psi_{\mathbf{n},1} \\ \psi_{\mathbf{n},2} \\ ...
\end{matrix}
\right)
\label{intspin}
\end{equation}
the dimension of the spinor depends on the representation $r$ of the gauge group used for it.
Such fields may have local (mass) terms in the Hamiltonian, with the most general form
\begin{equation}
H_M = \underset{\mathbf{n}}{\sum} M_{\mathbf{n}} \psi_{\mathbf{n}}^{\dagger}\psi_{\mathbf{n}}
\label{Hmass}
\end{equation}
where summation on the group indices is implicitly included, of course ($\psi_{\mathbf{n}}^{\dagger}\psi_{\mathbf{n}} = \underset{a}{\sum}\psi_{\mathbf{n},a}^{\dagger}\psi_{\mathbf{n},a}$ ). These terms are gauge invariant,
as the group indices are fully contracted. Another way to see the gauge invariance is to consider the explicit local gauge transformation on the matter fields,
\begin{equation}
\psi_{\mathbf{n}} \rightarrow V^{r}_{\mathbf{n}}\psi_{\mathbf{n}} = \underset{b}{\sum}\left(V^{r}_{\mathbf{n}}\right)_{ab}\psi_{\mathbf{n}b}
\label{psitrans}
\end{equation}
where $V_\mathbf{n}$ is an element of the group, represented by the unitary matrix $V_\mathbf{n}^r$ in the same representation of $\psi_{\mathbf{n}}$, defined locally for each vertex $\mathbf{n}$, and see that it leaves these terms
invariant.

The interactions among the matter fields must include the gauge fields as well, being the force mediators. As such, the most reasonable place for the gauge degrees of freedom is on the lattice's \emph{links}. Thus, on each link of
the lattice, emanating from the vertex $\mathbf{n}$ in direction $k$, define a group element $U_{\mathbf{n},k}$ (see figure \ref{lattice}). These elements can be represented by any representation $r$ of the group. In general, $U_{\mathbf{n},k}$ are matrices of operators, defined on the link's local Hilbert space. This matrix space is called "group space", and the matrix indices, $a,b$ etc., are referred to as "group indices". Explicit examples of such $U_{\mathbf{n},k}$ matrices of operators will be shortly presented.

\begin{figure}
  \centering
  \includegraphics[scale=1]{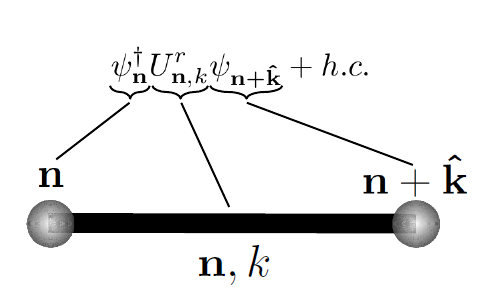}
  \caption{The elementary interactions - the interaction between the neighboring vertices $\mathbf{n}$ and $\mathbf{n + \hat k}$ involves the gauge field on the link $\mathbf{n},k$ connecting them: thus the gauge bosons are interaction
  mediators. As discussed in section \ref{elemint}, this is the "natural" type of interactions in our simulation scheme, included in the fundamental atomic Hamiltonian.}
  \label{linkint}
\end{figure}

The interaction between neighboring vertices is mediated using the link connecting them, in the form of \emph{elementary interactions} (see figure \ref{linkint})
\begin{equation}
\begin{aligned}
H_{int} = &\epsilon \underset{\mathbf{n},k}{\sum}\left(\psi_{\mathbf{n}}^{\dagger} U^r_{\mathbf{n},k}  \psi_{\mathbf{n+\hat{k}}} + h.c. \right) = \\
 & \epsilon \underset{\mathbf{n},k}{\sum}\underset{a,b}{\sum}\left(\psi_{\mathbf{n},a}^{\dagger} \left(U^r_{\mathbf{n},k}\right)_{ab}  \psi_{\mathbf{n+\hat{k}},b} + h.c. \right)
 \end{aligned}
 \label{fbhop}
\end{equation}
where $U^r$ is the matrix representation of $U$ in $r$, the spinors' representation. Once again, these terms are gauge invariant as all the group's indices are contracted. One can also deduce, from the transformation law of the
spinors (\ref{psitrans}) and the gauge invariance demand, the transformation law of the $U$'s,
\begin{equation}
\begin{aligned}
U^r_{\mathbf{n},k} & \rightarrow  V^{r}_{\mathbf{n}}U^r_{\mathbf{n},k}V^{\dagger r}_{\mathbf{n+\hat{k}}} \\
\left(U^r_{\mathbf{n},k}\right)_{ab} & \rightarrow  \underset{c,d}{\sum}\left(V^{r}_{\mathbf{n}}\right)_{ac}\left(U^r_{\mathbf{n},k}\right)_{cd}\left(V^{\dagger r}_{\mathbf{n+\hat{k}}}\right)_{db}
\label{utrans}
\end{aligned}
\end{equation}

Finally, one shall introduce the pure-gauge terms as well.
One type of pure-gauge terms which is gauge invariant, is of the form
\begin{equation}
H_E = \frac{g^2}{2} \underset{\mathbf{n},k,a}{\sum} \left(E_{\mathbf{n},k}\right)_a \left(E_{\mathbf{n},k}\right)_a
\label{HEgen}
\end{equation}
where $\left(E_{\mathbf{n},k}\right)_a$ are the generators of the group's algebra - for example, an angular momentum algebra for $SU(2)$, consisting of the angular momentum operators as generators. This term is just a sum of Casimir operators (which commute with all the generators, such as the total angular momentum operator for $SU(2)$), and it is interpreted as the "electric energy". In general, one can define left ($L_a$) and right ($R_a$) generators on each link, constrained to give the same Casimir operator
$\underset{a}{\sum}L_aL_a=\underset{a}{\sum}R_aR_a \equiv \underset{a}{\sum}E_aE_a$.

These electric field also construct the generators of local gauge transformations,
\begin{equation}
\left(G_{\mathbf{n}}\right)_a = \text{div}_{\mathbf{n}}E_a - Q_{\mathbf{n}}
\label{gausslaw}
\end{equation}
where $\text{div}_{\mathbf{n}}E_a$ is the "discrete divergence" of the group,
\begin{equation}
\text{div}_{\mathbf{n}}E_a = \underset{k}{\sum}\left(\left(L_{\mathbf{n}, k}\right)_a-\left(R_{\mathbf{n-\hat k},k}\right)_a\right)
\end{equation}
and $Q_{\mathbf{n}}$ is the local charge (either dynamic or static). These generators are constant of motion - the physical states are the gauge invariant ones,
satisfying, for each vertex $\mathbf{n}$, the Gauss's law
\begin{equation}
\left(G_{\mathbf{n}}\right)_a \left|phys\right\rangle = 0
\label{Gauss}
\end{equation}

\begin{figure}
  \centering
  \includegraphics[scale=1]{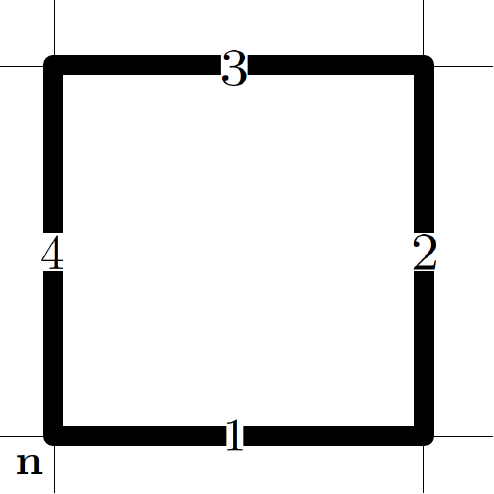}
  \caption{The plaquette interactions - the gauge-gauge interactions. The labeling of the links around the plaquette is according to equation (\ref{pla}). As discussed in section \ref{plaqs}, these interactions are obtained effectively in our
  simulation scheme.}
  \label{plaqint}
\end{figure}

Other type of gauge invariant Hamiltonian terms is the trace of group elements along a closed path. The shortest such paths are plaquettes, and they form the "magnetic energy" part (see figure \ref{plaqint}),
\begin{equation}
H_B = -\frac{1}{g^2} \underset{\text{plaquettes}}{\sum} \left(\text{Tr}\left(U_1 U_2 U^\dagger_3 U^\dagger_4\right) + h.c. \right)
\label{pla}
\end{equation}
where the $1,2,3,4$ links are oriented along a plaquette (see figure \ref{plaqint}). The trace is on group (matrix) indices.

Usually one includes in the Hamiltonian all such terms, where all the objects (group elements and spinors) are chosen to be in the fundamental representation. This will also be our choice throughout the paper.

Next, let us give some explicit examples of three gauge theories we use and simulate in this paper. We shall describe the structure and Hilbert space of these theories, whereas further details can be found in the appendix.

\subsubsection{Compact QED}
\label{cQED}
As a basic example, we discuss compact QED (cQED). This is an abelian gauge theory, with the gauge group $U(1)$, whose continuum limit is regular QED. However, unlike continuous QED, this theory manifests confinement of charges: at
all values of the coupling constant $g$ for $1+1$ and $2+1$ dimensions, and in the strong coupling regime for $3+1$ dimensions \cite{Polyakov,BanksMyersonKogut,DrellQuinnSvetitskyWeinstein,KogutLattice}. The compactness of the
theory is essential for the existence of a confining phase \cite{BenMenahem}.

In this theory, the $U_{\mathbf{n},k}$ operators defined on the links are pure phases: $U_{\mathbf{n},k} = e^{i \phi_{\mathbf{n},k}}$. The conjugate electric field $E_{\mathbf{n},k}$ is merely an angular momentum operator, taking
integer values from $-\infty$ to $\infty$. Thus, on each link $U_{\mathbf{n},k}$ the Hilbert space is the one of a quantum rotor, with canonical variables satisfying
\begin{equation}
\left[E_{\mathbf{n},k},\phi_{\mathbf{m},l}\right] = -i \delta_{\mathbf{nm}}\delta_{kl}
\end{equation}
this makes the $U$ operators ladder operators of angular momentum - or, in other words, of electric flux:
\begin{equation}
U_{\mathbf{n},k} \left|m\right\rangle= e^{ i \phi_{\mathbf{n},k}}\left|m\right\rangle = \left|m + 1 \right\rangle
\end{equation}
Note that as this group is abelian, there is no need to use different left and right generators. As there is only one generator, Gauss's law (\ref{Gauss}) simplifies to
\begin{equation}
\left(G_{\mathbf{n}}\right) \left|phys\right\rangle = 0
\label{aGauss}
\end{equation}
where $G_{\mathbf{n}} =  \underset{k}{\sum}\left(E_{\mathbf{n}, k}-E_{\mathbf{n-\hat k},k}\right)$.

\begin{figure}
  \centering
  \includegraphics[scale=1.5]{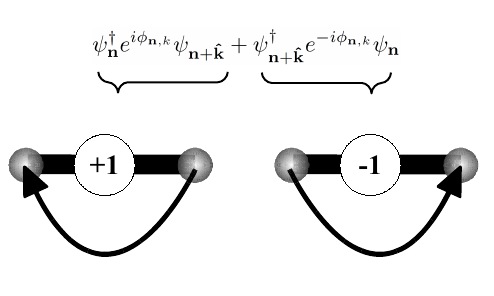}
  \caption{Illustration, using the $U(1)$, of the directionality of elementary interactions \ref{abhint}. If a fermion hops to the left, the flux in the middle increases. If a fermion hops to the right, the flux in the middle decreases.}
  \label{fluxjump}
\end{figure}

Using these operators, we can deduce from the general Hamiltonians (\ref{HEgen}),(\ref{pla}) the abelian version of the Kogut-Susskind Hamiltonian,
\begin{multline}
H_{KS} = H_E + H_B = \\
\frac{g^2}{2} \underset{\mathbf{n},k}{\sum} E_{\mathbf{n},k}^2 - \frac{1}{g^2}\underset{\mathbf{n}}{\sum}\cos\left(\phi_{\mathbf{n},1}+\phi_{\mathbf{n + \hat 1},2}-\phi_{\mathbf{n +  \hat 2},1}-\phi_{\mathbf{n},2}\right)
\label{aHKS}
\end{multline}
In the continuum limit, $H_E$ is identified with the electric energy and $H_B$ with the magnetic one (as the cosine's argument is the curl of the vector potential:
$\cos\left(\phi_{\mathbf{n},1}+\phi_{\mathbf{n + \hat 1},2}-\phi_{\mathbf{n +  \hat 2},1}-\phi_{\mathbf{n},2}\right) \rightarrow 1 - \frac{B^2}{2}$).
As for $H_{int}$, using staggered fermions \cite{Susskind1977} (see the appendix), we only have one spinor at each vertex, and in this case (\ref{fbhop}) is simplified to
\begin{equation}
H_{int} = \epsilon \underset{\mathbf{n},k}{\sum}\left(\psi_{\mathbf{n}}^{\dagger} e^{i \phi_{\mathbf{n},k}}  \psi_{\mathbf{n+\hat{k}}} + \psi_{\mathbf{n+\hat{k}}}^{\dagger} e^{-i \phi_{\mathbf{n},k}}  \psi_{\mathbf{n}} \right)
\label{abhint}
\end{equation}
and thus, the basic interaction involves a fermion hopping between neighboring vertices, while raising/lowering, \emph{depending on the direction of the hopping fermion}, the electric flux on the link connecting them
(see figure \ref{fluxjump}).

For further details, refer to the appendix.

\subsubsection{$\mathbb{Z}_N$ gauge theory}
\label{ZN}
Here we shall review the properties of a Hamiltonian $\mathbb{Z}_N$ gauge theory \cite{Horn1979}. We restrict ourselves to the pure gauge case, as only this is relevant for the purposes of this paper.

First, let us describe the local Hilbert space on every link of the lattice.
Define two operators, $P,Q$, which are unitary:
\begin{equation}
P^{\dagger}P = Q^{\dagger}Q = 1
\end{equation}
and satisfy the $\mathbb{Z}_N$ algebra,
\begin{equation}
P^N = Q^N =1 \quad;\quad P^{\dagger}QP = e^{i \delta}Q
\end{equation}
where $\delta = \frac{2\pi}{N}$.

For example, one can work with the basis of $P$ eigenstates,
\begin{equation}
P \left|m\right\rangle = e^{i m \delta}\left|m\right\rangle
\end{equation}
with $m \in \left\{ -N/2, ... ,N/2\right\}$ (without loss of generality, we assume $N$ is odd - the change for an even $N$ is straightforward)
and then $Q$ is a unitary ladder operator,
\begin{equation}
Q \left|m\right\rangle = \left|m-1\right\rangle
\end{equation}
with the cyclic property $Q \left|-N/2\right\rangle = \left|N/2\right\rangle$.
Altenatively, one can expand the Hilbert space in terms of $Q$ eigenstates, and then $P$ will be a unitary raising operator (with the cyclic property, again).

Interestingly, one can introduce the Hermitean operators $E,A$ on every link, by
\begin{equation}
P = e^{i \delta E} ; Q = e^{i A}
\end{equation}
and then, for $N \rightarrow \infty$, one obtains the cQED Hilbert space, with canonically conjucate $E,A$.

Let us now combine the entire lattice in order to get the gauge invariant Hamiltonian. It has the form
\begin{multline}
H = H_E + H_B = \\
-\frac{1}{2} \mu \underset{\mathbf{n},k}{\sum}\left(P_{\mathbf{n},k} + P^{\dagger}_{\mathbf{n},k}\right) \\
 - \frac{1}{2}\underset{\mathbf{n}}{\sum}\left( Q_{\mathbf{n},1} Q_{\mathbf{n + \hat 1},2} Q^{\dagger}_{\mathbf{n + \hat 2},1} Q^{\dagger}_{\mathbf{n},2} + h.c.\right)
\end{multline}

One can define a static \emph{modular} charge on the vertex ${\mathbf{n}}$, by $q_{\mathbf{n}} = e^{-i \delta m}$. Then the Gauss's law means, that a gauge invariant state must satisfy (for every ${\mathbf{n}}$),
\begin{equation}
G_{\mathbf{n}}\left|phys\right\rangle = q_{\mathbf{n}}\left|phys\right\rangle
\end{equation}
where
\begin{equation}
G_{\mathbf{n}} = \underset{l+}{\prod}P_{l+}^{\dagger}\underset{l-}{\prod}P_{l-} = e^{-i \delta \text{div}_{\mathbf{n}}E}
\end{equation}
with $l_+$ are links starting at $\mathbf{n}$ (positive links), and $l-$ are ending there (negative links).

Since these charges are modular and thus very different than the charges of continuous gauge theories, we shall only consider the pure-gauge case for $\mathbb{Z}_N$ in this paper.

\subsubsection{$SU(N)$ gauge theories}
\label{SUN}
Let us first discuss the links' Hilbert space for a pure gauge theory. In a representation
$r$, with representation matrices $\left\{ T_{a}^{r}\right\} $,
the group elements can be parametrized as
\begin{equation}
U^{r}_{\mathbf{n},k}=e^{i \underset{a}{\sum} T_{a}^{r}\phi_{\mathbf{n},k}^a}
\end{equation}
- it is a matrix in group space.

Due to the non-abelian nature of the gauge group, it must have separate
left and right generators, $\left\{ L_{a}\right\} ,\left\{ R_{a}\right\} $
respectively, corresponding to {}``left and right'' non abelian
electric fields \cite{KogutSusskind,Kogut1983}:
they can be represented as differential operators, canonically conjugate
to the group parameters $\left\{ \phi_{\mathbf{n},k}^a\right\} $.
As left and right generators of the group, they obey the following
commutation relations with the group elements (within the same link,
of course)
\begin{equation}
\left[L_{a},U^{r}\right]=T_{a}^{r}U^{r}\quad;\quad\left[R_{a},U^{r}\right]=U^{r}T_{a}^{r}\label{eq:groupcom}
\end{equation}
and the Lie algebra
\begin{equation}
\left[L_{a},L_{b}\right]=-if_{abc}L_{c}\quad;\quad\left[R_{a},R_{b}\right]=if_{abc}R_{c}
\label{SU2algebra}
\end{equation}
where $f_{abc}$ are the group's structure constants
\footnote{One should note that one can have $\left[L_{a},L_{b}\right]=if_{abc}L_{c}$.
That results in a redefinition of the left generators, resulting in
sign changes in its commutator in (\ref{eq:groupcom}), gauge generators
(\ref{gausslaw}) and Gauss's law, etc.}, and also $\left[L_{a,}R_{b}\right]=0$. Physically, the difference
between the left and right generators of a link may be interpreted
as the \emph{color charge} of it. The left and right generators can
be obtained from each other using the group element on the link in
the adjoint representation.

From the local gauge transformation (\ref{utrans}), one can conclude that the generators of local
groups transformation are (\ref{gausslaw}) - where in the pure-gauge case $Q_{\mathbf{n}}$ are C-numbers.

\begin{figure}
\centering{}\includegraphics{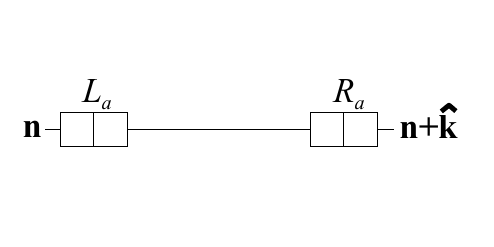}\caption{The link's operators. Left and right generators $\left\{ L_{a}\right\} ,\left\{ R_{a}\right\} $,
and the group element. The generators transform in the adjoint representation
$\left(j=1\right)$, as symbolized by the Young tableaux, and are
related to each other by the rotation matrix $U^{1}\left(\mathbf{n},k\right)$
(which is in the adjoint representation as well).}
\label{su2link}
\end{figure}

From now on, we shall focus mostly on $SU\left(2\right)$, the
simplest continuous non-abelian group. There \cite{KogutSusskind},
$r=j$ (total angular momentum quantum number), $f_{abc}=\epsilon_{abc}$.
The local Hilbert space is characterized by three integer quantum
numbers, $j,m,m'$, which are eigenvalues of the Casimir operators
and the $z$ components of left and right angular momentum:
\begin{multline}
\underset{a}{\sum}E_{a}E_{a}\left|jmm'\right\rangle= \underset{a}{\sum}L_{a}L_{a}\left|jmm'\right\rangle =\\
\underset{a}{\sum}R_{a}R_{a}\left|jmm'\right\rangle =j\left(j+1\right)\left|jmm'\right\rangle
\end{multline}
\begin{equation}
L_{z}\left|jmm'\right\rangle =m\left|jmm'\right\rangle \quad;\quad R_{z}\left|jmm'\right\rangle =m'\left|jmm'\right\rangle
\end{equation}
the link Hilbert space may be interpreted as the one of a rigid rotator.
The generators in the two edges of a link may then be interpreted
as generators of rotations in the body/space systems \cite{KogutSusskind}.
The link's structure, in terms of operators and Young tableaux, is
presented in figure \ref{su2link}.

What shall be the Hamiltonian of such a theory? If we require it to
be gauge-invariant, it may contain only gauge-invariant terms. Such
terms can be constructed out of the generators, and since they must
be contracted we get the Casimir operators $L_{a}L_{a}=R_{a}R_{a}$.
They construct the local part of the Hamiltonian, called the \emph{Electric
Part} - $H_E$ (\ref{HEgen}). Other possibilities are closed loops: the "most local"
ones are the traces of group elements directed around a single plaquette,
forming the \emph{Magnetic Part}. We choose them to be in the fundamental
representation ($T_{a}=\frac{1}{2}\sigma_{a}$, where $\sigma_{a}$
are Pauli matrices) to get $H_B$ (\ref{pla}), and finally obtain the $SU\left(2\right)$ version
of the Kogut-Susskind Hamiltonian \cite{KogutSusskind}, $H=H_E+H_B$.

\subsection{Basic requirements for a HEP quantum simulation}
\label{RHEP}
As can be understood from above, quantum simulation of HEP may be of great interest, and also significance, as it may help in avoiding problems of classical simulation, such as the sign problem. However, one should note it requires
much more complex ingredients than quantum simulation of condensed matter systems - quantum simulation of HEP models must:
\begin{enumerate}
    \item{Include \emph{both fermions and bosons}, if one wishes to simulate both matter and gauge fields. This requires, for a cold-atom simulations, the use of many different atomic species.}
    \item{Respect \emph{local gauge invariance}, in order to have the correct symmetry which is responsible to the interactions and the interesting special features of the theories.}
    \item{Be \emph{relativistic}. This can be reduced, if a lattice gauge theory is simulated, demanding that the continuum limit will still be relativistic.}
\end{enumerate}

If one chooses to work on the lattice, as we do, the local gauge invariance "problem" transforms to the challenge of obtaining two types of interactions. First, the "link" gauge-matter interaction (\ref{fbhop}), which couples the
matter and gauge field degrees of freedom in a very special way;
Our basic idea is to get these interactions fundamentally in the atomic Hamiltonian -
they will be derived directly from the conservation of hyperfine angular momentum $\mathbf{F}$ in atomic collisions.

The second type of interactions is the plaquette interactions (\ref{pla}), which are, essentially, four-body interactions - not a fundamental part of the atomic Hamiltonian. However, as we show, these terms can be obtained
effectively from the link terms, using perturbation theory \cite{Soliverez1969,CTBook}. Although they are obtained effectively, gauge invariance is still fundamental, as the building blocks - elementary interactions, already fulfill the gauge symmetry.

\subsection{HEP toy models}
\label{toy}
Simulations of gauge theories, which must satisfy all the three requirements presented above, are challenging. However, when simulating HEP phenomena, one would not necessarily need, in first stage, to simulate the entire standard
model, or even Quantum Chromodynamics. Several simpler models are available for observing the important phenomena and phases of the complicated theories. For example, working on the lattice, Compact QED is suitable
for observing confinement (see appendix 1): although everyday continuous QED manifests the opposite behavior of a Coulomb phase, the compact lattice theory contains a confining phase in the strong coupling limit of the $3+1$ dimensional theory, and
confines for any value of the coupling constant in the $1+1$ and $2+1$ dimensional theories \cite{Polyakov,BanksMyersonKogut,DrellQuinnSvetitskyWeinstein,KogutLattice}. Thus, for the observation of confinement in a pure gauge theory,
simulation of $2+1$ cQED is enough (the $1+1$ dimensional model is trivial). If one wishes to introduce dynamic charges, even the $1+1$-d dimensional case is interesting - for example, one could simulate the lattice version of the
Schwinger model \cite{Banks1976}.

As for non-abelian theories, full-fledged QCD with an $SU(3)$ gauge symmetry is not essentially required as well, for the first step. A lot of theoretical, both qualitative and quantitative insight has been gained on QCD using the
$1+1$-dimensional version of the theory, QCD$_2$, or more generally, $SU(N)$ in $1+1$ dimensions \cite{tHooft1974,Callan1976,Witten1984,Frishman1993,Hornbostel1990,Gross1996,Armoni2001}. On the other hand, some phenomena, such as
confinement, may be observed also using a smaller gauge group - $SU(2)$. Thus, for simulations of non-abelian gauge theories, $SU(2)$ on the lattice \cite{Hamer1977}, even in $1+1$ dimensions, is enough.

\subsection{Summary of previous works}
\label{prevsum}

 Several suggestions have been made for simulations of quantum field theories which do not include gauge fields. These include the observation of vacuum entanglement of a scalar field using trapped ions \cite{Retzker2005}, and the
 simulation of interacting scalar and fermionic fields - Thirring and Gross-Neveu models using cold atoms \cite{Cirac2010} (the latter could also be interpreted as a 1+1 simulation of fermions coupled to a gauge field).
These two models correspond to simulation of fields in the continuum, respecting the appropriate relativistic and causal structure. Quantum computation of scattering amplitudes for scalar field theories was introduced in \cite{Jordan2012,Jordan2011}.
Simulations of fermionic lattice QFTs have been proposed as well, where the fermions were either free or in external non-dynamical gauge fields. These include Axions and Wilson fermions \cite{Bermudez2010}, Dirac fermions in curved
spacetime \cite{Boada2011} and general quantum simulators of QFTs and topological insulators \cite{Mazza2012}.

As for abelian pure gauge theories, simulation of $2+1$-d cQED, with the possibility to observe confinement, first using Bose-Einstein condensates (BECs) of ultracold atoms in optical lattices \cite{Zohar2011} and then with single
atoms in optical lattices \cite{Zohar2012} have been suggested, where the first is of the abelian Kogut-Susskind Hamiltonian \cite{KogutSusskind} and the latter of a truncated "Spin-Gauge" theory.

The inclusion of dynamical matter in such theories of great interest as well. This was done either for the link model \cite{Horn1981,Orland1990,Wiese1997,Brower1999} - a $1+1$-d simulation of the lattice Schwinger model \cite{Banerjee2012},
or as a generalization of the previous pure-gauge simulations in $2+1$-d, to include dynamical fermions \cite{Zohar2013}. The latter also suggested a way to realize the gedanken experiment proposed in \cite{Topological} of measuring
Wilson-Loop's area law.

All these abelian proposals fulfilled the relativity requirement through the use of the lattice. The models which included simulations fulfilled the first requirement by either including both fermions and bosons, or enabling the
simulation of both types of particles. The gauge invariance demand has also been met, however, it has not been done in a direct way: gauge symmetry is not fundamental in these models, but rather appears as a low-energy symmtery,
manifested in the dynamics of an effective Hamiltonian, obtained using a Gauss's law constraint required to introduce gauge invariance: this is since the four-body plaquette interactions are not fundamental for optical lattices.
In \cite{Higgs}, the possibility of interpreting the breaking of the Gauss's law constraint as the emergence of Higgs fields is discussed, in the context of the simulation proposed in \cite{Zohar2011}.

Simulations of other abelian lattice gauge theories are \cite{Szirmai2011,Tagliacozzo2013}. One should also note the continuum QED simulation proposed in \cite{Kapit2011} (which does not manifest confinement as it is not
compact \cite{BenMenahem}).

 Some proposals for the simulations of non-abelian models have already been proposed as well, either utilizing prepotentials \cite{MathurSU2}, using ultracold atoms in optical lattices \cite{NA} or utilizing Rishons in the
 link model \cite{Rishon2012}. In both methods a constraint is used in order to obtain the desired interactions. A digital simulation of an $SU(2)$ gauge magnet \cite{Orland1990,Orland2012} has also been suggested \cite{TagliacozzoNA}.
 In \cite{Rishon2012}, as in non-abelian link models, the original symmetry is larger and one has to break it in order to obtain the right symmetry group; In \cite{NA}, the $SU(2)$ gauge symmetry is \emph{fundamental} and is
 manifested already by the basic atomic Hamiltonian, unlike the effective methods of the former abelian simulations: this is done by exploiting the fundamental angular momentum conservation of the atoms, in a way which will be
 further explained and utilized for simulating other gauge theories in the next sections of this paper.

 A realization of discrete gauge theories (such as $\mathbb{Z}_N$) has been discussed using Josephson junctions \cite{Doucot2004}.

\subsection{The present work}
\label{present}
We have introduced the requirements from the quantum simulations of a gauge theory. As explained, in the previous proposals for simulations of abelian theories, gauge invariance was effective, rather than exact. Here we shall
describe the way to utilize a fundamental symmetry in systems of ultracold atoms in order to get a gauge symmetry which is not an effective low-energy symmetry, but rather built-into the theory, and thus is more robust.

\begin{figure}
  \centering
  \includegraphics[scale=1]{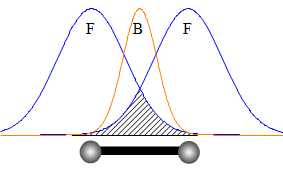}
  \caption{A schematic plot of the overlap of the fermionic Wannier functions (F) of two neighboring vertices and the bosonic Wannier functions (B) on the link. This is since the bosonic overlap is of order 1 and the fermionic tunneling is thus maximal.}
  \label{overlap}
\end{figure}

In the simulating scheme we suggest in this paper, we use fermions as matter and bosons as gauge fields, in vertices
and links, respectively, like in the previous proposals. However,
\begin{enumerate}
    \item{ \emph{Local Gauge invariance.} We do not impose Gauge invariance using an energy penalty in the Hamiltonian.
            Instead, we show that by a judicious choice of fermionic and bosonic
            species (i.e., internal states), the natural atomic scattering interactions give
            rise to the terms we need with the appropriate gauge symmetries. This
            is so because the gauge symmetry in the resulting HEP model is equivalent
            to the angular momentum conservation in the collisions in the atomic model.
            }
    \item{ \emph{Elementary interactions on links.} The interaction terms between bosons and fermions are chosen so that
           they are maximal, and can compete with the real tunneling. This is obtained
           by using the idea of fig. \ref{overlap} (see next section).
     \item{ \emph{Plaquette interactions.} In $2+1$ dimensional systems (and more), in order to obtain the dynamical terms of the gauge bosons
           (plaquette terms), we must use fourth order perturbation theory, introducing the loop method. This would naively mean that
           we get very small terms. However, we make sure
           that the odd orders are cancelled (or just renormalize previous terms), so that
           in reality the conditions are equivalent to a second order perturbation theory, that are not so small.
           The plaquette interactions are $O(\epsilon ^4)$ (where $\epsilon$ is defined in eq. (\ref{fbhop})),
           but $\epsilon, \epsilon^3$ and the other odd orders of $\epsilon$ are absent in the perturbative series,
           and thus the expansion parameter is $\epsilon^2$ and effectively it is a second order contribution: $O\left(\left(\epsilon ^2\right)^2\right)$.
           }

    }
\end{enumerate}

       Resulting from that, we
\begin{enumerate}
        \item{Propose a 1+1 dimensional cQED simulation, which should be relatively simple to implement experimentally.}
        \item{Extend it to 2+1 dimensional cQED by adding plaquette terms, and also introduce a new $2+1$ dimensional model, $\mathbb{Z}_N$.}
        \item{Suggest a method for simulation of $SU(N)$ theories, including a possible extension of the $SU(2)$ model considered in \cite{NA}  to $2+1$ dimensions.}
\end{enumerate}

\section{The simulating system}
\label{simsys}

\begin{figure}
  \centering
  \includegraphics[scale=1.5]{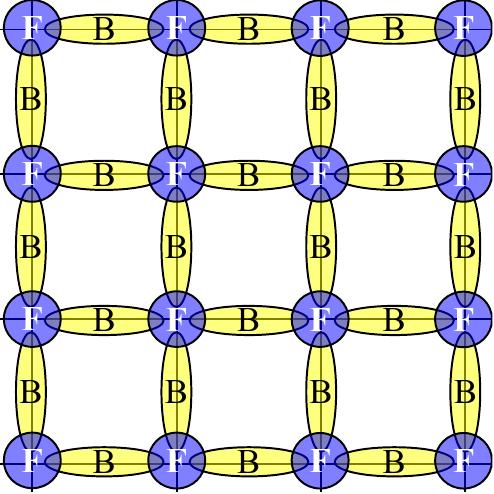}
  \caption{Schematic structure of the optical lattice used for simulations: Bosonic minima on the links (B), and fermionic minima on the vertices (F).}
  \label{optlat}
\end{figure}

Let us consider the atomic ingredients. We would like to build a theory of both fermions and bosons, where the fermions reside on the vertices, and the bosons - on the links (see figure \ref{optlat}). Thus, let us start
with the most general such structure. The vertices $\mathbf{n}$ of a square optical lattice coincide with the minima of fermions, described by the second-quantization operators $\Psi_{\alpha}\left(\mathbf{x}\right)$, where $\alpha$
labels the atomic species. Each link of this lattice coincides with a bosonic minimum, in which the bosons $\Phi_{\alpha}\left(\mathbf{x}\right)$ may reside.
If one assumes that the single-particle energy levels of each minimum are remote enough, only the lowest Bloch bands may be considered, and thus the second-quantized field operators may be expanded in terms of local annihilation
operators $c_{\mathbf{n},\alpha},a_{\mathbf{n},k,\alpha}$ and local Wannier functions $\psi_{\mathbf{n},\alpha}\left(\mathbf{x}\right),\phi_{\mathbf{n},k,\alpha}\left(\mathbf{x}\right)$, for fermions on the vertex $\mathbf{n}$
and bosons on the link emanating from it to the $\mathbf{\hat k}$ direction respectively:
\begin{equation}
    \begin{aligned}
        \Psi_{\alpha}\left(\mathbf{x}\right) = & \underset{\mathbf{n},\alpha}{\sum}c_{\mathbf{n},\alpha}\psi_{\mathbf{n},\alpha}\left(\mathbf{x}\right) \\
        \Phi_{\alpha}\left(\mathbf{x}\right) = & \underset{\mathbf{n},k,\alpha}{\sum}a_{\mathbf{n},k,\alpha}\phi_{\mathbf{n},k,\alpha}\left(\mathbf{x}\right)
    \end{aligned}
\label{aexp}
\end{equation}

The most general atomic Hamiltonian contains the following terms:
\begin{enumerate}

\item{Single particle terms:
    \begin{equation}
    H_0=\underset{\alpha}{\sum}\int d^3\mathbf{x} \left(\Psi^{\dagger}_{\alpha}\left(\mathbf{x}\right) H_{0,f} \Psi_{\alpha}\left(\mathbf{x}\right) +
    \Phi^{\dagger}_{\alpha}\left(\mathbf{x}\right) H_{0,b} \Phi_{\alpha}\left(\mathbf{x}\right) \right)
    \end{equation}
    where $H_0,f$, $H_0,b$ are the single particle Hamiltonians, containing the kinetic energy and the trapping potentials, for the fermions and bosons respectively.
    Once the expansion (\ref{aexp}) is plugged into these terms, and the overlap of Wannier functions is taken into account in the integration, one obtains two types of terms: local terms, linear in the atomic numbers, and
    nearest-neighbor hopping terms. In order to eliminate the latter for bosons, one should design the bosonic lattice deep enough such that any interactions outside a bosonic minimum would be negligible; In order to avoid fermionic
    tunneling, one could use different species at neighboring vertices, alternately.
    }

\item{Scattering terms:
    \begin{multline}
    H_{sc} = \underset{\alpha,\beta,\gamma,\delta}{\sum}g^{FF}_{\alpha \beta \gamma \delta}\int d^3\mathbf{x}
    \Psi^{\dagger}_{\alpha}\left(\mathbf{x}\right)\Psi^{\dagger}_{\beta}\left(\mathbf{x}\right) \Psi_{\gamma}\left(\mathbf{x}\right) \Psi_{\delta}\left(\mathbf{x}\right) + \\
    \underset{\alpha,\beta,\gamma,\delta}{\sum}g^{BB}_{\alpha \beta \gamma \delta}\int d^3\mathbf{x}
    \Phi^{\dagger}_{\alpha}\left(\mathbf{x}\right)\Phi^{\dagger}_{\beta}\left(\mathbf{x}\right) \Phi_{\gamma}\left(\mathbf{x}\right) \Phi_{\delta}\left(\mathbf{x}\right) + \\
    \underset{\alpha,\beta,\gamma,\delta}{\sum}g^{BF}_{\alpha \beta \gamma \delta}\int d^3\mathbf{x}
    \Psi^{\dagger}_{\alpha}\left(\mathbf{x}\right)\Psi_{\beta}\left(\mathbf{x}\right) \Phi^{\dagger}_{\gamma}\left(\mathbf{x}\right) \Phi_{\delta}\left(\mathbf{x}\right)
    \label{Hsc}
    \end{multline}
    where the scattering coefficients $g^{FF}_{\alpha \beta \gamma \delta},g^{BB}_{\alpha \beta \gamma \delta},g^{BF}_{\alpha \beta \gamma \delta}$
    are constrained by conservation laws and are fixed for different atoms - but can be controlled and modified using Feshbach resonances
    (perhaps optical, \cite{Fedichev1996,Bohn1997,Fatemi2000}, if more than one is required). The first two terms represent the fermion-fermion and boson-boson scattering. Their integration, using (\ref{aexp}),
    yields local scattering terms, within the same minima.

    }

\item{Rabi (laser) terms:
    \begin{equation}
    \begin{aligned}
    H_R = & \underset{\alpha,\beta}{\sum} \Omega^F_{\alpha \beta} \int d^3\mathbf{x} \Psi^{\dagger}_{\alpha}\left(\mathbf{x}\right)\Psi_{\beta}\left(\mathbf{x}\right)  \\
    + &\underset{\alpha,\beta}{\sum} \Omega^B_{\alpha \beta} \int d^3\mathbf{x} \Phi^{\dagger}_{\alpha}\left(\mathbf{x}\right)\Phi_{\beta}\left(\mathbf{x}\right)
    \end{aligned}
    \end{equation}
    Using such terms, one can create "manually" desired hopping processes which may be useful in several cases.
    }
\end{enumerate}

 In principle one could also consider molecular terms, which dissociate into atoms, giving rise to a term with a bosonic operator and two fermionic ones. This may be useful for simulations in the bulk, but will not be used in the present paper where we concentrate on lattices.

\section{Elementary interactions along links}
\label{elemint}

 In this section we show how the fermion-gauge boson interaction terms appear in a natural way in the atomic system if one makes a judicious choice of internal states. Since the elementary interactions must come from the scattering of fermions with bosons (eq. (\ref{Hsc})), it must involve an overlap integral between the initial and final bosonic and fermionic states. Given the fact that the fermions must hop, those two states will be located at different positions. In order to make this term as large as possible, one must have the bosonic atoms placed in between the fermionic ones (see figure \ref{overlap}). Furthermore, in order to satisfy the gauge symmetry we will choose that the fermions and bosons change the internal states in this process according to the angular momentum conservation.

The key idea is \emph{angular momentum conservation}: in these atomic scattering processes, the total hyperfine angular momentum $\mathbf{F}_{tot}$ is conserved.
In particular, the $z$ components - $m_F$, are conserved.  One can specifically select the $m_F$ values of the atomic species utilized, in order to generate the required interactions over the link, and eliminate the others.
    This will
    result in only gauge invariant terms, and forms the correspondence between two fundamental symmetries: angular momentum conservation in the atomic, simulating level, is equivalent to gauge invariance in the simulated

Let us first discuss the case of an abelian $H_{int}$, as in (\ref{abhint}).

\subsection{$U(1)$ elementary interactions}
\label{Uel}

For simulating cQED, we need two fermionic species and two bosonic species \footnote{As will be later explained, one could use only one fermionic species, and
replace the fermions with more bosons.}, arranged in an optical lattice, as in the previous section. Let us first consider a one dimensional lattice, and thus the links may be labeled only by one index - the vertex from
which they emanate - however, the same method may be generalized for more spatial dimensions, as we shall later do.

We start with the bosons. Denote that bosonic species $a,b$, both having two different value of $m_F$. As explained before, no interactions take place between bosons of different links. Thus the total number of bosons on
each link is a constant of motion - we denote it by $N_0$, setting it equal all around the lattice, and taking it to be an even number.

On each link, a Schwinger algebra \cite{Jordan,Schwinger} is constructed from the two bosonic species:
\begin{equation}
L_+ = a^\dagger b ; L_- = b^\dagger a
\end{equation}
and
\begin{equation}
L_z = \frac{1}{2}\left(N_a - N_b\right) ; \ell = \frac{1}{2}\left(N_a + N_b\right) = \frac{N_0}{2}
\end{equation}
where $L_z$ is our (truncated) electric field.

Next, we wish to consider the fermions. As we would like to simulate a staggered fermions model (\cite{Banks1976,Susskind1977}, see appendix 2), we only need a single fermion at most on each vertex. However, in order to
use angular momentum conservation to ensure gauge invariance, we must use two different fermions, labeled by $c$ and $d$, arranged such that the $c$ minima occur in even vertices and the $d$ minima in odd vertices. This eliminates the fermionic
nearest-neighbor tunneling of $H_0$. The lattice is designed such that we get from $H_0$ the mass Hamiltonian
\begin{equation}
H_M = M \underset{n}{\sum}(-1)^n\psi^{\dagger}_{n}\psi_{n}
\label{MQED}
\end{equation}
where $\psi_{n}$ is either $c_n$ or $d_n$, depending on the parity of the vertex. The Dirac sea state is obtained if initially all the $d$ vertices are filled and the $c$ vertices are empty. Note that if $M>0$, the
fermionic minima do not have to be $m_F$-dependent: if the system is initially prepared in a gauge invariant state, the fermionic tunneling of $H_0$ is energetically forbidden and thus effectively
eliminated and can be disregarded. Moreover, it also assures that two fermions can never occupy a single vertex (even of a different species), and thus the fermion-fermion scattering terms of $H_{sc}$ may be disregarded.
The fermions' local charges are defined as
$Q_{n} = \psi_{n}^{\dagger}\psi_{n} - \frac{1}{2}\left(1-\left(-1\right)^{n}\right)$ - for further details, see appendix 2.

\begin{figure}
  \centering
  \includegraphics[scale=1.5]{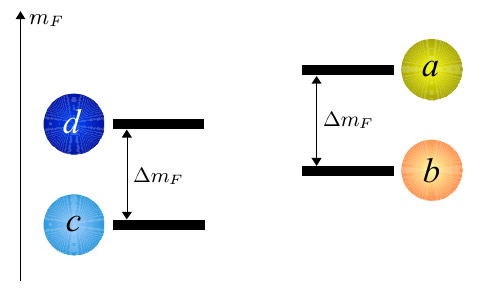}
  \caption{A schematic plot of the required choice of $m_F$ values. As in equation (\ref{eqmfs}), the equal spacing between $\Delta m_F$ the $m_F$ of bosons ($a,b$ on the right) and fermions ($c,d$ on the left) is required
  to allow only "gauge invariant" collisions.}
  \label{mfs}
\end{figure}

\begin{figure}
  \centering
  \includegraphics[scale=1.5]{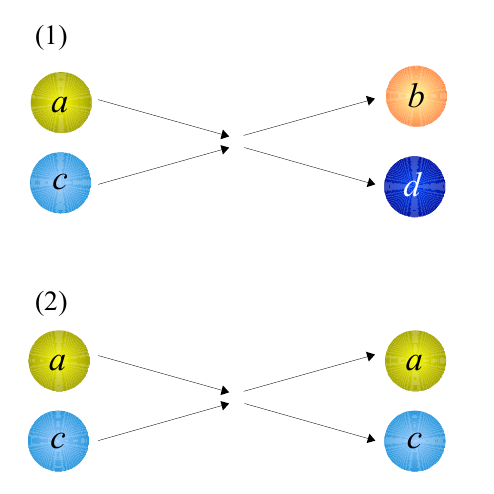}
  \caption{Schematic examples of the two types of possible Boson-Fermion scattering processes (collisions) along a link: (1) species changing and (2) non species changing. Both of them conserve total $m_F$ and thus are gauge
  invariant.}
  \label{colls}
\end{figure}

The gauge invariant elementary interactions are obtained from the boson-fermion scattering (the third part of $H_{sc}$ (\ref{Hsc})). This is done by utilizing the total $m_F$ in atomic collisions.
The hyperfine levels of the participating atoms should satisfy
\begin{equation}
m_F \left(a\right) + m_F \left(c\right) = m_F \left(b\right) + m_F \left(d\right)
\label{eqmfs}
\end{equation}
(see figure \ref{mfs}) and thus, the only $m_F$ conserving processes (collisions) are (see figure \ref{colls}):
\begin{enumerate}
    \item{$a,c \rightarrow b,d$, and vice versa. This yields terms like $c_{n}^{\dagger}a_{n}^{\dagger}b_{n}d_{n+1} + d_{n+1}^{\dagger}b_{n+1}^{\dagger}a_{n+1}c_{n+2}$}
    \item{$a,c \rightarrow a,c$, and the same with $b,d$. This results in terms like $c_{n}^{\dagger}c_{n}\left(a_{k}^{\dagger}a_{k}+b_{k}^{\dagger}b_{k}\right)$ where the link $k$ starts or ends in the vertex $n$. As all
    this terms $a_{k}^{\dagger}a_{k}+b_{k}^{\dagger}b_{k} = N_0$, one eventually gets a contribution which is proportional to the constant number of fermions - an ignorable constant in the energy.}
\end{enumerate}

The first term is the desired gauge-invariant interaction. To see that, just perform the canonical transformation
\begin{equation}
\left(\begin{matrix}
    a_{n} \\ b_n
\end{matrix}\right)
\rightarrow
\sigma_x^n
\left(\begin{matrix}
    a_{n} \\ b_n
\end{matrix}\right)
\label{cantra}
\end{equation}
and redefine the scattering coefficients in $H_{sc}$, to obtain
\footnote{One can also introduce different phases to $H_{int}$, as is sometimes done, by using another canonical transformation, of the form $\psi_n \rightarrow \left(-i\right)^n \psi_n$. \label{fphase}}
\begin{equation}
H_{int} = \frac{\epsilon}{\sqrt{\ell \left(\ell + 1\right)}} \underset{n}{\sum}\left(\psi_n^{\dagger}L_{+,n}\psi_{n+1} + h.c. \right)
\label{HintS}
\end{equation}

This Hamiltonian is especially interesting (although not realizable) in the limit $N_0 \rightarrow \infty$. In that case, $\ell \rightarrow \infty$, and thus always $m \ll \ell$. Thus $L_{\pm}$, in this limit, are unitary operators:
\begin{equation}
\frac{L_{\pm}}{\sqrt{\ell \left(\ell + 1\right)}}\left|\ell m \right\rangle =
\sqrt{1 - \frac{m\left(m \pm 1\right)}{\ell \left(\ell + 1\right)}}\left|\ell, m \pm 1 \right\rangle \underset{\ell \rightarrow \infty}{\longrightarrow}\left|\ell, m \pm 1 \right\rangle
\end{equation}
and thus we get that $\frac{L_{\pm}}{\sqrt{\ell \left(\ell + 1\right)}}$ approaches in this limit a unitary operator (pure phase), as in the Kogut-Susskind model.
Another way to see it, is to consider that in this case the bosons form BECs. For $N_0 \gg 1$, one can approximate $a_n \approx \sqrt{\frac{N_0}{2}} e^{-i \theta^a_n} = \sqrt{\ell} e^{-i \theta^a_n}$ etc.
($m \ll \ell$ and thus $N_a - N_b \ll N_0$ and it is reasonable to approximate $N_a \approx \frac{N_0}{2}, N_b \approx \frac{N_0}{2}$ Then,
\begin{equation}
\frac{1}{\sqrt{\ell \left(\ell + 1\right)}}\psi_{n}^{\dagger}a_{n}^{\dagger}b_{n}\psi_{n+1} \approx \psi_{n}^{\dagger}e^{i \phi_n}\psi_{n+1}
\end{equation}
where $\phi_n = \theta^a_n - \theta^b_n$. This is similar to the mapping of \cite{Zohar2011}.

\subsection{$\mathbb{Z}_N$ elementary interactions}
\label{Zel}
Now we turn to the construction of the elementary interactions of another abelian LGT theory, but this time with a \emph{discrete} gauge group - $\mathbb{Z}_N$.
Besides being an interesting gauge theory on its own, we consider its quantum simulation due to the fact that here, in order to simulate the \emph{exact} theory, with \emph{exactly unitary} gauge operators in the elementary intercations,
we need a finite number of degrees of freedom, which makes this theory more tempting to realize, unlike the $cQED$ case, in which we only approximated the unitary interactions by angular momentum ladder operators.

Although we use the same general techniques of angular momentum conservation, one must note that in this case it is not enough. This is due to the fact that the $\mathbb{Z}_N$ $Q$ operators are cyclic (see section \ref{ZN} details),
forming an "Escher's staircase" \cite{Mueller2004}, and thus regular angular momentum conservation is not sufficient. Therefore we use on top of the angular momentum conservation hybridization of states, and make use of auxiliary bosonic levels.

For simplicity, we describe here the construction of elementary interactions of the $\mathbb{Z}_3$ case in $2+1$ dimensions, but it can be easily generalized for larger $N$s and higher dimensions. We do not consider the $1+1$ dimensional
theory as our fermions are not $\mathbb{Z}_3$ charges, but rather auxiliary particles which shall be traced out in the derivation of plaquette interactions (see section \ref{Zpl}): this point will become clear in the following derivation
of the elementary interactions.

\begin{figure}
  \centering
  \includegraphics[scale=1.5]{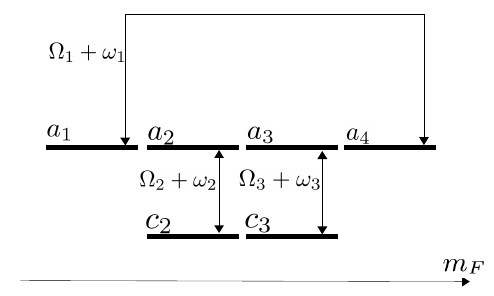}
  \caption{A schematic plot of the required choice of $m_F$ values, for the four "regular" levels $\left\{a_i\right\}_{i=1}^4$, and the two "auxiliary" levels $\left\{c_i\right\}_{i=2}^3$ as in equation (\ref{zms}).
  Note that this plot shows $m_F$ only - not all the processes are available, only the "gauge invariant" ones, as in (\ref{HintZ}):
  some processes are eliminated due to energy shifts, which are not drawn in this figure. Also schematically shown (by arrows) are the lasers connecting coupled of levels in $H_R$ (\ref{RRR})}
  \label{Znmfs}
\end{figure}

For obtaining the elementary interactions of $\mathbb{Z}_3$, we need, on each link, six fermionic species: four "regular" hyperfine levels, which we label $\left\{a_i\right\}_{i=1}^4$, and two "auxiliary" levels $\left\{c_i\right\}_{i=2}^3$
(see figure \ref{Znmfs}). The vertices, unlike before, are occupied by bosons, whose annihilation operators are $\psi_{\mathbf{n}} ,\chi_{\mathbf{n}}$ which, due to energy shifts, can occupy alternating vertices (like the
fermions in the Schwinger model simulation, and thus the use of "fermionic" letters).
The vertex bosons are subject to a hard-core constraint,
\begin{equation}
H_c = \lambda \underset{v}{\sum}N_v\left(N_v-1\right)
\end{equation}
where $N_v$ is the total number of bosons on the vertex $v$. Initially all the vertices are filled by only one boson - even vertices with $\psi$ and odd with $\chi$.

 The hyperfine levels of the atoms should satisfy the relation
 \begin{equation}
 m_f \left(\psi\right) + m_f \left(a_i\right) = m_f \left(\chi\right) + m_f \left(a_{i+1}\right)
 \label{zms}
 \end{equation}
 for $i \in \left\{1,2,3\right\}$ (see figure \ref{Znmfs}). The $c$ levels should be picked far enough energetically, such that they will not be involved in any link-species-changing process. Thus, the boson-fermion
 scattering terms will be of the the two following forms:
 \begin{enumerate}
    \item{Collisions with no change of species:
        \begin{equation}
        H_{\alpha} = \alpha \underset{\left\langle l,v \right\rangle}{\sum}N_v N_l
        \end{equation}
        where $\left\langle l,v \right\rangle$ are neighboring links and vertices, $N_v$ is the total number of fermions on the vertex $v$ and $N_l$ - the total number of bosons on the link $l$. We prepare the system
        initially with $N_l =1$. This is not changed by any interaction, and thus this term turns out to be proportional to the total number of fermions in the system - an ignorable constant.
    }
    \item{Collisions with a change of species. For even vertices (emanating from an even $\mathbf{n}$), we have
        \begin{equation}
        2\epsilon \underset{k}{\sum}\left(\psi^{\dagger}_{\mathbf{n}} \underset{i=1}{\overset{3}{\sum}} a^{\dagger}_{i,\mathbf{n},k}a_{i,\mathbf{n},k} \chi_{\mathbf{n + \hat k}} +h.c \right)
        \end{equation}
        For odd links, one has to replace $\psi \leftrightarrow \chi$ in the equation above. However, we can perform a canonical transformation, "inverting" the names of hyperfine levels on odd links, and then have them described
        by the same sort of interaction - compare to the canonical transformation of (\ref{cantra})). After doing that we call, formally, all the vertex bosons $\psi$ and obtain
        \begin{equation}
        H_{int} = 2\epsilon \underset{\mathbf{n},k}{\sum}\left(\psi^{\dagger}_{\mathbf{n}} \underset{i=1}{\overset{3}{\sum}} a^{\dagger}_{i,\mathbf{n},k}a_{i,\mathbf{n},k} \psi_{\mathbf{n + \hat k}} +h.c \right)
        \label{HintZ}
        \end{equation}
    }
 \end{enumerate}
 Since there is only one boson on each link, other boson-boson scattering processes are irrelevant.

 We also introduce, using Raman lasers (see figure \ref{Znmfs}), for each link, the following bosonic tunneling Hamiltonian, within each link:
 \begin{multline}
 H_R = \left(\Delta_1 + \delta_1\right)\left(a^{\dagger}_1 a_1 + a^{\dagger}_4 a_4\right) + \left(\Omega_1 + \omega_1\right)\left(a^{\dagger}_1 a_4 + a^{\dagger}_4 a_1\right) \\
+  \left(\Delta_2 + \delta_2\right)\left(a^{\dagger}_2 a_2 + c^{\dagger}_2 c_2\right) + \left(\Omega_2 + \omega_2\right)\left(a^{\dagger}_2 c_2 + c^{\dagger}_2 a_2\right)  \\
+ \left(\Delta_3 + \delta_3\right)\left(a^{\dagger}_3 a_3 + c^{\dagger}_3 c_3\right) + \left(\Omega_3 + \omega_3\right)\left(a^{\dagger}_3 c_3 + c^{\dagger}_3 a_3\right)
 \label{RRR}
 \end{multline}
Note that the $a_1 \leftrightarrow a_4$ process involves a three-unit angular momentum change, and thus it should be mediated by three photons.

 We make $\Delta_i, \Omega_i$  the largest energy scales in the total Hamiltonian, and thus it will be reasonable to digaonalize $H_R$ first, and obtain a hybridization of the couples of states coupled with lasers,
 in the form of a Bogolyubov transformation:
 \begin{equation}
 \begin{aligned}
 b_1^{\dagger} = \frac{1}{\sqrt{2}}\left(a_1^{\dagger} + a_4^{\dagger}\right) & ; & d_1^{\dagger} = \frac{1}{\sqrt{2}}\left(a_1^{\dagger} - a_4^{\dagger}\right)\\
 b_2^{\dagger} = \frac{1}{\sqrt{2}}\left(a_2^{\dagger} + c_2^{\dagger}\right) & ; & d_2^{\dagger} = \frac{1}{\sqrt{2}}\left(a_2^{\dagger} - c_2^{\dagger}\right)\\
 b_3^{\dagger} = \frac{1}{\sqrt{2}}\left(a_3^{\dagger} + c_3^{\dagger}\right) & ; & d_3^{\dagger} = \frac{1}{\sqrt{2}}\left(a_3^{\dagger} - c_3^{\dagger}\right)
\end{aligned}
 \label{Bog}
 \end{equation}

 By setting $\Omega_i = -\Delta_i$, we obtain the diagonalized form
 \begin{equation}
 H_R = 2 \underset{i,\mathbf{n},k}{\sum}\Delta_i d^{\dagger}_{i,\mathbf{n},k}d_{i,\mathbf{n},k} +  \left(\delta,\omega \text{ terms}\right)
 \end{equation}
 and since we choose $\Delta$ to be very large, we can disregard, effectively, the $d_i$ modes. Plugging the Bogolyubov transformation (\ref{Bog}) into $H_{int}$, disregarding the $d$ modes, we get
 \begin{equation}
 H_{int} = \epsilon \underset{\mathbf{n},k}{\sum}\left(\psi^{\dagger}_{\mathbf{n}} Q_{\mathbf{n},k} \psi_{\mathbf{n + \hat k}} +   h.c. \right)
 \end{equation}
 where
 \begin{equation}
 Q = b_1^{\dagger}b_2 + b_2^{\dagger}b_3 + b_3^{\dagger}b_1
 \end{equation}
 is the unitary $Q$ of $\mathbb{Z}_3$ (see section \ref{ZN} for details).
Thus $H_{int}$ is the desired $\mathbb{Z}_3$ elementary interaction. Note, however, that the vertex bosons do not represent $\mathbb{Z}_3$ charges, and thus this method can only be used to generated auxiliary particles,
and not dynamic charges.

Finally, we have to represent the electric part, $H_E$. Plugging the new modes into $H_R$, we get
\begin{equation}
H_R = \underset{i,\mathbf{n},k}{\sum}\left(\left(\delta_i + \omega_i\right)b^{\dagger}_{i,\mathbf{n},k}b_{i,\mathbf{n},k} + \left(\delta_i - \omega_i\right)d^{\dagger}_{i,\mathbf{n},k}d_{i,\mathbf{n},k}\right)
\end{equation}
Then, setting $\delta_i = \omega_i$, one gets
\begin{equation}
H_R = 2\underset{i,\mathbf{n},k}{\sum}\delta_i b^{\dagger}_{i,\mathbf{n},k}b_{i,\mathbf{n},k}
\end{equation}

We identify
\begin{equation}
P + P^{\dagger} = 2\underset{m}{\sum}\cos\left(m \delta\right) b^{\dagger}_m b_m
\end{equation}
(where $\delta = 2\pi / N$), and thus, for $\mathbb{Z}_3$,
\begin{equation}
H_E = \frac{\mu}{2} \underset{\mathbf{n},k}{\sum}\left(b^{\dagger}_{1,\mathbf{n},k}b_{1,\mathbf{n},k} -2b^{\dagger}_{2,\mathbf{n},k}b_{2,\mathbf{n},k} + b^{\dagger}_{3,\mathbf{n},k}b_{3,\mathbf{n},k} \right)
\end{equation}
Set $\delta_1 = \delta_3 = \frac{\mu}{2}$ and $\delta_2 = - \mu$ and obtain, neglecting constants,
\begin{equation}
H = H_C + H_R + H_{int} = H_C + H_E + H_{int}
\label{HfundZ}
\end{equation}
- this is the fundamental Hamiltonian, with \emph{unitary} elementary interactions, from which we can now construct effectively the $\mathbb{Z}_3$ Hamiltonian with plaquette terms.

Note that in order to obtain the correct interactions, one must use several Feshbach resonances. Their number can be reduced, if one generalizes the Hamiltonian to include some energy difference
between $a_1^{\dagger}a_1$ and $a_4^{\dagger}a_4$, and also for the other two coupled of hybridized states. Then one can introduce a few more parameters to play with, and reduce the number of required Feshbach resonances.

Also, note that in order to generalize to $\mathbb{Z}_N$ for $N>3$, one must have $2N$ bosonic species on each link, $\left\{a_i\right\}_{i=1}^{N+1}$ and $\left\{c_i\right\}_{i=2}^{N}$. The hybridization method is
the same, with coupling between $a_1$ and $a_N$, and $a_i$ and $c_i$ for $i \in \left\{2,...,N\right\}$.

\subsection{$SU(N)$ Yang Mills elementary interactions}
\label{Sel}

\begin{figure}
  \centering
  \includegraphics[scale=1.5]{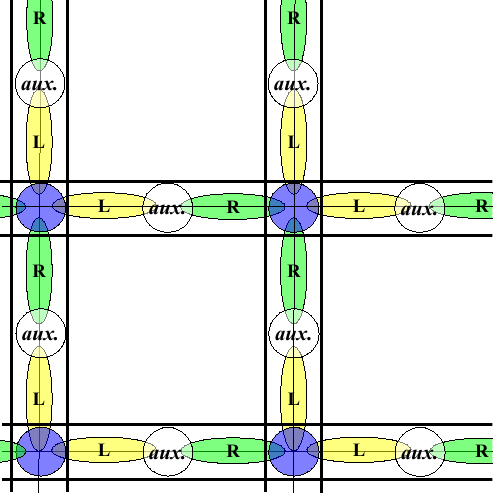}
  \caption{The lattice structure required for elementary $SU(N)$ interactions. Each link is decomposed to left (L) and right (R) parts. It is constructed from two links of the optical lattice,
  tailored together by a constrained satisfied by auxiliary (aux) fermions in the middle.}
  \label{SUnlattice}
\end{figure}

We shall also refer to the fundamental Hamiltonian for $SU(N)$ elementary interactions.
There, the system is more complicated, and many atomic species are required. Due to the decomposition of a single link to two parts (left and right, see section \ref{SUN}), this richer Hilbert space requires the construction of a single
link of what was two separate links for the abelian theories - i.e., a simulating link is effective, and it is constructed from two atomic links, tailored by some constrained auxiliary fermions between them (see figure
\ref{SUnlattice}).

We shall briefly review the ideas of \cite{NA}, in which a non-abelian quantum simulator for a $1+1$ dimensional $SU(2)$ lattice gauge theory was suggested.
This simulation method utilizes prepotentials \cite{MathurSU2}, in which the group degrees of freedom are constructed out of "prepotentials" - harmonic oscillators, or, in our case, bosonic species. This enables a bosonic representation of the full Kogut-Susskind model. Fermionic representations are available too, using the link model \cite{Wiese1997,Brower1999} - however, they correspond to truncated gauge theories, with finite local Hilbert spaces, from which one obtains the full theories only in the continuum limit.

As explained in \ref{SUN}, and in figure \ref{SUnlattice}, each link is decomposed into two parts, the left and the right, and hence simulated by two links of the optical lattice. In each of the link's parts, four bosonic species reside: $a_1,a_2,c_1,c_2$ on the left, and $b_1,b_2,d_1,d_2$ on the right. The $a,b$ species are the gauge field degrees of freedom, forming, using a Schwinger representation, the left and right generators of the group, respectively,
\begin{equation}
L_a = \frac{1}{2}\underset{k,l}{\sum}a_k^{\dagger}\left(\sigma_a\right)_{lk}a_l
 \; ; \;
R_a = \frac{1}{2}\underset{k,l}{\sum}b_k^{\dagger}\left(\sigma_a\right)_{kl}b_l
\label{angmom}
\end{equation}
satisfying the required algebra of the group (eq. \ref{SU2algebra}), with $j = \frac{N_L}{2} = \frac{N_R}{2}$ and the Casimir operators $\mathbf{L}^2 = \frac{N_L}{2}\left(\frac{N_L}{2} + 1\right)$, $\mathbf{R}^2 = \frac{N_R}{2}\left(\frac{N_R}{2} + 1\right)$ (where $N_L \equiv a_1^{\dagger}a_1 + a_2^{\dagger}a_2$ and $N_R \equiv b_1^{\dagger}b_1 + b_2^{\dagger}b_2$, satisfying the constraint $N_L = N_R$).

From these, in the prepotential method, one may construct the left and right matrices (of operators), in the fundamental representation,
\begin{equation}
U_L = \frac{1}{\sqrt{N_L + 1}}\left(
        \begin{array}{cc}
          a_1^{\dagger} & -a_2 \\
          a_2^{\dagger} & a_1 \\
        \end{array}
      \right)
 \; ; \;
U_R = \left(
        \begin{array}{cc}
          b_1^{\dagger} & b_2^{\dagger} \\
          -b_2 & b_1 \\
        \end{array}
      \right)
      \frac{1}{\sqrt{N_R + 1}}
\end{equation}
and obtain the group element on the link, in the fundamental representation,
\begin{equation}
U = U_L U_R
\end{equation}
satisfying the required commutation relations (\ref{eq:groupcom}).

The $c,d$ species are prepared in coherent states (Bose-Einstein condensate) $\left|\alpha\right\rangle$, where $\alpha \in \mathbb{R}, \alpha \gg 1$.

Let us denote the "real" spinors by $\psi$ and the auxiliary ones, connecting between two links which will form one link in the simulated theory, $\chi$. Then, by properly choosing the hyperfine levels of all the atoms (see the supplemental material of \cite{NA} for an explicit example), and tuning the scattering coefficients, we get the angular-momentum conserving interaction Hamiltonian for elementary interactions,
\begin{multline}
H_{int} = \frac{\epsilon}{2^{1/4}\alpha} \underset{n,i,j}{\sum} \Bigg(\left(\psi_n^{\dagger}\right)_i\left(\tilde W_{L,n}\right)_{ij} \left(\chi_n\right)_j + \\ + \left(\chi_n^{\dagger}\right)_i\left(\tilde W_{R,n}\right)_{ij} \left(\psi_{n+1}\right)_j + h.c. \Bigg)
\label{ints}
\end{multline}
where
\begin{equation}
\tilde W_L = \left(
        \begin{array}{cc}
          a_1^{\dagger}c_1 & -a_2c^{\dagger}_2 \\
          a_2^{\dagger}c_2 & a_1c^{\dagger}_1 \\
        \end{array}
      \right)
 \; ; \;
\tilde W_R = \left(
        \begin{array}{cc}
          b_1^{\dagger}d_1 & b_2^{\dagger}d_2 \\
          -b_2d^{\dagger}_2 & b_1d^{\dagger}_1 \\
        \end{array}
      \right)
\end{equation}
and we label the two links from which the effective link $n$ (emanating from the "real" vertex $n$) will be generated by $n,L$ and $n,R$.

The use of condensates for the auxiliary bosonic species allow us to replace $c_i,d_i$ by $\alpha$, and since $\alpha \gg 1$ we can approximately do the same for $c_i^{\dagger},d_i^{\dagger}$, and one effectively obtains the Hamiltonian
\begin{multline}
H_f = \frac{\epsilon}{2^{1/4}} \underset{n,i,j}{\sum} \Bigg(\sqrt{N_{L,n}+1}\left(\psi_n^{\dagger}\right)_i\left(U_{L,n}\right)_{ij} \left(\chi_n\right)_j + \\ + \left(\chi_n^{\dagger}\right)_i\left(U_{R,n}\right)_{ij} \left(\psi_{n+1}\right)_j\sqrt{N_{R,n}+1} + h.c. \Bigg)
\end{multline}

The auxiliary fermions are constrained by the large-scale energy constraint
\begin{equation}
H_{\chi} = \lambda\underset{n}{\sum}\chi^{\dagger}_n \chi_n
\end{equation}
If initially the system does not contain any $\chi$ fermions, and $\lambda$ is the largest energy scale, we can obtain, using second order perturbation theory, an effective Hamiltonian, tailoring the two sides of each link, of the form
\begin{equation}
H_{int}^{eff} = \frac{\epsilon_{eff}}{\sqrt{2}} \underset{n}{\sum}\left(\psi_{n}^{\dagger}\sqrt{N_{L,n}+1}U_n\sqrt{N_{R,n}+1}\psi_{n+1} + h.c. \right)
\label{Hinteff}
\end{equation}

These are the elementary interactions of $SU(2)$. However, note that the bosonic link operators are not unitary, i.e. we have $\sqrt{N_{L,n}+1}U_n\sqrt{N_{R,n}+1}$ rather than $U_n$. In spite of that, as will be explained in the next section, one can still get qualitatively the same physics, in the appropriate parameter regime.

Also note, that although the full link is obtained effectively, the gauge invariance is still fundamental and it is constructed out of two already gauge-invariant building blocks: the left and right parts.

The prepotentials method can of course be generalized to $SU(N)$ gauge theories with $N>2$ \cite{Mathur2006,Mathur2007,Anishetty}, and serve as a base for obtaining the elementary interactions in a similar manner.

\section{$1+1$ dimensional models}
\label{1p1}
Having the elementary interactions in hand, we can now construct complete quantum simulations of $1+1$ dimensional gauge theories with dynamic fermions. This can be done for $cQED$, but not for $\mathbb{Z}_N$, as we do not discuss discrete charges in this paper. A proposal based on our method for the simulation of $1+1$ dimensional $SU(2)$ theory has already been suggested in \cite{NA}, and we shall review it here as well.

\subsection{Quantum Simulation of the Schwinger model}
\label{Sch}
Let us start with the a quantum simulation of the Schwinger model: a $1+1$ dimensional abelian gauge theory (QED) coupled to dynamical fermions (see appendix 2).
The solvable Schwinger model involves \emph{massless} fermions. We discuss also the more general massive case. Being $1+1$ dimensional, this
system does not involve any plaquette interactions, and thus we already have all the interactions we need.

Besides $H_{int} + H_M$ (eqs. (\ref{MQED}),(\ref{HintS})), we also need the electric Hamiltonian. $H_E = \frac{g^2}{2}\underset{n,k}{\sum}E_{n}^2 = \frac{g^2}{2}\underset{n}{\sum}L_{z,n}^2$, or, in the atomic terms,
\begin{equation}
H_E = \frac{g^2}{8}\underset{n}{\sum}\left(N_{a,n}^2+N_{b,n}^2 -2N_{a,n}N_{b,n}\right)
\end{equation}
This is exactly obtained from the boson-boson scattering terms of (\ref{Hsc}). These processes, of course, conserve the total $m_F$ in collisions. The minus sign in the interaction may be avoided as well, thanks to the
constant $N_0$: one could, instead, use the Hamiltonian
\begin{equation}
\begin{aligned}
H'_E = & \left(\alpha + \frac{g^2}{8}\right)\underset{n}{\sum}\left(N_{a,n}^2+N_{b,n}^2\right)  \\
+ & \left(2\alpha - \frac{g^2}{4}\right)\underset{n}{\sum}N_{a,n}N_{b,n}
\end{aligned}
\end{equation}
which is just $H_E$, plus a constant in the energy - $\alpha\underset{n}{\sum}\left(N_{a,n}+N_{b,n}\right) = \alpha\underset{n}{\sum}N_0^2$ - which is, of course, ignorable.

Linear terms in the total number of bosons on a link (from $H_{sc}$ and $H_0$) yield ignorable constants as well.

Thus we get the Hamiltonian
\begin{equation}
H = H_E + H_M + H_{int}
\label{HSchwinger}
\end{equation}
describing the dynamics of a $U(1)$ "Spin-Gauge" theory with dynamic fermions \cite{Zohar2012,Zohar2013} in $1+1$-dimensions.

For a finite $N_0$, one gets qualitatively the features of the model. As $N_0$ (or $\ell$) increases, the model becomes more accurate. The phase approximation can be made for condensates,
in which one must make sure that three-body interactions are negligible. This can be assumed if the condensate is made in the shape of a tube, whose axis is perpendicular to the link,
increasing the number of particles but reducing their density.

Thus we have shown how to simulate a $1+1$-d cQED with dynamic staggered fermions (lattice Schwinger model) using ultracold atoms, with an \emph{exact gauge symmetry} and \emph{no use of perturbation theory}
and effective low-energy considerations unlike in previous suggestions.

\subsection{Quantum simulation of $1+1$ dimensional $SU(2)$ gauge theory}
\label{SU1}

Having also the $SU(2)$ elementary interactions in hand, one can obtain a $1+1$ dimensional simulation of an $SU(2)$ gauge theory.

On top of the elementary interactions (\ref{Hinteff}), one shall include as well the electric and matter Hamiltonians,
\begin{equation}
H_E = \frac{1}{2} \underset{n}{\sum}\left(g_L\frac{N_{L,n}}{2}\left(\frac{N_{L,n}}{2} + 1\right)+g_R\frac{N_{R,n}}{2}\left(\frac{N_{R,n}}{2} + 1\right)\right)
\label{PGH}
\end{equation}
with $g_R + g_L = g^2$,
and
\begin{equation}
H_M = M\underset{n}{\sum}(-1)^{n} \psi^{\dagger}_n \psi_n
\end{equation}

This enables a simulation of the dynamics of the vacuum of the theory, up to fifth order perturbation theory in $H_{int}^{eff}$ (\ref{Hinteff}). See \cite{NA} and its supplemental material for further details.

\section{Interactions on plaquettes: The loop method}
\label{plaqs}

In the next step, we would like to generalize our discussion to further dimensions. However, the $1+1-d \rightarrow 2+1-d$ transition is nontrivial, since the plaquette terms must be introduced, and, as explained,
they are not a fundamental part of the atomic Hamiltonian. In previous proposals, the plaquette terms have been obtained effectively, by constraining the Gauss's law, and introducing gauge invariance effectively,
 as a symmetry of the low-energy sector. In this section, we show yet another way to get the plaquette terms. Although in what we shall describe the plaquettes will be obtained effectively as well, it is believed
 to be much more robust than the previous methods, since although we get the plaquette terms effectively, gauge invariance is fundamental as described in the previous sections: the building blocks, which are
 elementary interactions, are already gauge invariant. This is called "the loop method".

The idea is as follows. First, extend the scheme for $1+1$ simulations to more dimensions (which already serves as a simulation for the extreme strong limit). Then, treat the fermions as \emph{auxiliary} particles,
by adding a constraint which forces them to occupy only certain vertices (exactly one vertex belonging to each plaquette of the lattice), $H_C$ instead of $H_M$ (the auxilliary particles do not have to be massive). $H_{int}$, operating on states satisfying this constraint, will take us out
of the "right" sector. Thus it would be reasonable to construct an effective theory in the ground sector of this constrained Hamiltonian, and then, in fourth order (operating with elements of $H_{int}$ around each
plaquette) one obtains the required interaction. Remarkably, these fourth order terms are unnecassarily weak: as it turns out - and will be clarified throughout the following derivations - the relevant leading order is either the
second order (for abelian theories, with no third order) or the fourth one (for non abelian theories), and thus the perturbative parameter should be small only to order 2 or 4.

The nature of auxiliary particles varies from one gauge theory to another, and depends on the gauge group. We shall describe, seperately, the methods of constructing such simulations for three different gauge
theories: $U(1)$, $\mathbb{Z}_N$ and $SU(N)$. For the sake of simplicity, we consider the $2+1$-d case. However, at least by geometric means, the constructions for higher dimensions are similar.

As a final general remark, before getting into specific theories, one should note that in this method the fermions are "traced out" and eventually a pure-gauge theory is obtained. Such theories, in dimensions higher
than $1+1$, are interesting on their own, without including dynamic fermions; However, one could also include \emph{more} fermionic species, not subjected to the "plaquette constraint" described above, which will
serve as dynamical matter. This is discussed in section \ref{dynferm}.

\subsection{cQED plaquettes}
\label{Upl}

Our first example of effective derivation of plaquettes in the loop method will be for the case of cQED - a generalization of the Schwinger model simulation described in the previous section. For that, we start with a similar system
to the one described for $1+1$-d, but with two spatial dimensions instead of one. Thus, the system is described by the Hamiltonian (\ref{HSchwinger}), generalized to two dimensions: i.e., all the vertex indices
become vectors $n \rightarrow \mathbf{n} \in \mathbb{Z}^2$ and the links are now identified by two indices: $\mathbf{n}$ and the direction $k \in \left\{1,2\right\}$. The electric part of the Hamiltonian is
\begin{equation}
H_E = \frac{g^2}{2}\underset{\mathbf{n},k}{\sum}\left(E_{\mathbf{n},k}\right)^2
\end{equation}

\begin{figure}
  \centering
  \includegraphics[scale=1]{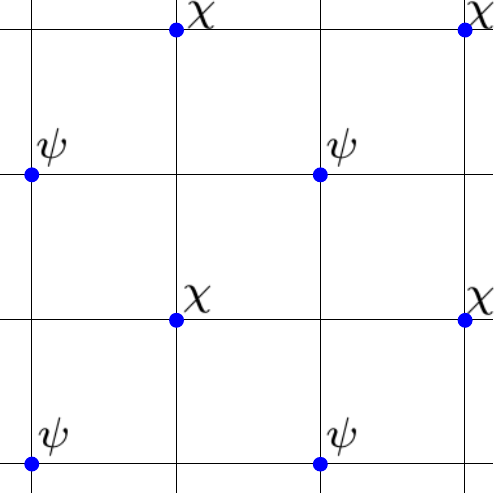}
  \caption{The "preferred" vertices of the $\psi,\chi$ fermions, in the even vertices, according to the constraint set by $H_C$ (equation (\ref{Hc})). Each plaquette contains exactly such two vertices, one of $\psi$ and one of $\chi$.}
  \label{spv}
\end{figure}

We introduce another set of fermions, $\chi_\mathbf{n}$, which behave like the $\psi_\mathbf{n}$s (including the same interactions with the bosons). At this stage, for the sake of illustration of the method for
obtaining plaquettes, we assume that the bosonic link operators are \emph{really} unitary ($N_0 \rightarrow \infty$), i.e. we work with the elementary interactions
\begin{multline}
H_{int} = \epsilon \underset{\mathbf{n},k}{\sum}\left(\psi_{\mathbf{n}}^{\dagger} U_{\mathbf{n},k}  \psi_{\mathbf{n+\hat{k}}} + \chi_{\mathbf{n}}^{\dagger} U_{\mathbf{n},k}  \chi_{\mathbf{n+\hat{k}}} + h.c.\right) =
\\ \epsilon \underset{\mathbf{n},k}{\sum}\left(\psi_{\mathbf{n}}^{\dagger} e^{i \phi_{\mathbf{n},k}}  \psi_{\mathbf{n+\hat{k}}} + \chi_{\mathbf{n}}^{\dagger} e^{i \phi_{\mathbf{n},k}}  \chi_{\mathbf{n+\hat{k}}}  + h.c.\right)
\end{multline}
Note that the hyperfine levels of the two $\chi$ species must be chosen carefully, such that $\chi-\psi$ remain \emph{separate} in $H_{int}$ and no mixing interactions can occur.
The difference between the types of fermions is found in a constraint we add on the fermions,
\begin{equation}
H_C = -\lambda \underset{\mathbf{n}}{\sum} \left(F_{\psi}\left(\mathbf{n}\right)  \psi_{\mathbf{n}}^{\dagger}\psi_{\mathbf{n}} + F_{\chi}\left(\mathbf{n}\right)\chi_{\mathbf{n}}^{\dagger}\chi_{\mathbf{n}}\right)
\label{Hc}
\end{equation}
where $F_{\psi}$ is zero everywhere, unless where both the indices $n_1,n_2$ are even, where it takes the value of 1, and $F_{\chi}$ is zero everywhere, unless where both the indices $n_1,n_2$ are odd, where it is 1.
If we define both these types
of vertices as \emph{even} ones, we see that $H_C$ puts an "energy penalty" for each species not being in its specific "preferred" type of an even vertex (see figure \ref{spv}).
$H_M$ is of course unnecessary here, as explained in the introduction of this section.

We denote the ground sector of $H_C$ as $\mathcal{M}_0$, and wish to work in this subspace. Thus, the system has to be initially prepared in a state where all the fermions occupy only even vertices (the opposite to
the Dirac sea case for dynamic fermions). Note that since $H$ is gauge invariant, and we initially prepare the system in a gauge invariant state, the dynamics will leave the state gauge invariant and thus we choose
to specifically work in
$\mathcal{M} \subset \mathcal{M}_0$, which is the set of gauge invariant state inside $\mathcal{M_0}$. As $\lambda$ is the largest energy scale, we derive an effective Hamiltonian within $\mathcal{M}_0$ - we shall
construct a low-energy effective theory which includes the plaquette interactions. In order to do that we use time-independent perturbation theory, following the notations of \cite{Soliverez1969}.

\subsubsection{First and second order contributions}
Denote $\mathcal{P}_0$ as the projection operator to $\mathcal{M}_0$, and define $H_1 = H_E + H_{int}$ and
\begin{equation}
\mathcal{K} = \underset{\left|\alpha\right\rangle \notin \mathcal{M}_0}{\sum}\frac{\left|\alpha\right\rangle\left\langle\alpha\right|}{E_C \left(\alpha\right) - E_C \left(0\right)}
\end{equation}
where $E_C$ is the eigenvalue of $H_C$, and thus $E_C \left(0\right)=0$. Also, for the convenience of series expansions, denote $\mu \equiv \frac{g^2}{2}$.

\begin{figure}
  \centering
  \includegraphics[scale=1]{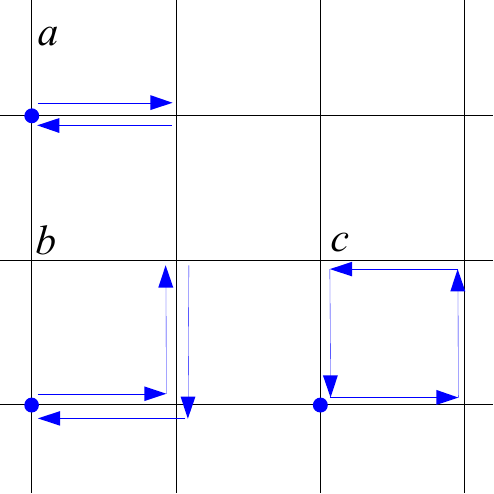}
  \caption{Examples for "jumping" of auxiliary fermions from their "preferred" vertices and back, (a) in the second order of $H_{int}$, (b,c) in the fourth order of $H_{int}$, where (c) forms the plaquette interactions.}
  \label{jumps}
\end{figure}

The first order term in the effective expansion is $H_{eff}^{\left(1\right)} = \mathcal{P}_0 H_1\mathcal{P}_0 = H_E$. In second order, we have $H_{eff}^{\left(2\right)} = -\mathcal{P}_0 H_1\mathcal{K} H_1\mathcal{P}_0$.
Here, in $\mathcal{K}$, only $H_1$ will contribute, taking to (and from) intermediate states $\left|\alpha\right\rangle$ with $E_C \left(\alpha\right) = \lambda$ (the constraint is violated only for one fermion).
The contributions will be only of double operations of $H_1$ on the same link (see figure \ref{jumps}a), and due to the unitarity of the interactions will lead (within $\mathcal{M}_0$) to a constant in the energy,
which is ignorable.

\subsubsection{Third order contributions}
\label{unit1third}
The third order contribution takes the form
$H_{eff}^{\left(3\right)} = \mathcal{P}_0 H_1\mathcal{K} H_1 \mathcal{K} H_1\mathcal{P}_0 - \frac{1}{2}\left\{\mathcal{P}_0 H_1\mathcal{K}^2 H_1\mathcal{P}_0,\mathcal{P}_0 H_1\mathcal{P}_0\right\}$. The second
(anti-commutator) term is just a combination of the first and second order terms, which will result in $-\mathcal{N}\frac{\epsilon^2}{\lambda^2}H_E$, where $\mathcal{N}$ is the number of links.

The first term is nonzero only for the combination $\mathcal{P}_0 H_{int}\mathcal{K} H_E\mathcal{P}_0\mathcal{K} H_{int}\mathcal{P}_0$. Here, as in the second order, $E_C \left(\alpha\right) = \lambda$. These
terms will vanish, unless we consider, as in the second order, double operation of $H_{int}$ on the same link. Define a "positive" link if it starts on an even vertex, and a "negative" link if it ends there.
Since only states where even vertices are occupied belong to $\mathcal{M}_0$, only the part $\psi_{\mathbf{n+\hat{k}}}^{\dagger} U^{\dagger}_{\mathbf{n},k}  \psi_{\mathbf{n}}$ of $H_{int}$
acting on $\mathcal{M}_0$ will give rise to a nonzero contribution, and thus in the final operation, the contribution will come from $\psi_{\mathbf{n}}^{\dagger} U_{\mathbf{n},k}  \psi_{\mathbf{n+\hat{k}}}$ (and
similarly for $\chi$). For negative links, only the opposite processes contribute. Thus, for each positive link $\mathbf{n},k$ we get the contribution
\begin{equation}
\frac{\epsilon^2}{\lambda^2} U H_E U^{\dagger} = \frac{\epsilon^2}{\lambda^2}H_E - \frac{2\mu \epsilon^2}{\lambda^2} E_{\mathbf{n},k} + \mathcal{D}
\end{equation}
where $\mathcal{C}$ is the same constant as in the other third order contribution, and  $\mathcal{D}$ is another constant. Similarly, for negative links we get
\begin{equation}
\frac{\epsilon^2}{\lambda^2} U^{\dagger} H_E U = \frac{\epsilon^2}{\lambda^2}H_E + \frac{2\mu \epsilon^2}{\lambda^2} E_{\mathbf{n},k} + \mathcal{D}
\end{equation}The $\mathcal{D}$ parts can be ignored, as constants, and after summing on all the links (remember that each link is a neighbor of exactly one even vertex, and thus it is either positive or negative)
we get a cancellation of the $\frac{\epsilon^2}{\lambda^2}H_E$ parts, and eventually we are left, for every even $\mathbf{n}$, with
$-\frac{2\mu \epsilon^2}{\lambda^2} \text{div}_{\mathbf{n}} E$, where the discrete divergence is $\text{div}_{\mathbf{n}} E = \left(E_{\mathbf{n},1}+E_{\mathbf{n},2}-E_{\mathbf{n-\hat1},1}-E_{\mathbf{n-\hat2},2}\right) = \text{const.}$
within $\mathcal{M}_0$ by Gauss's law (in case of no dynamic charges). Thus all the third order contributions are constants and we disregard them.

\subsubsection{Fourth order contributions}
In fourth order, there are much more contibutions. The ones involving $H_E$ yield mostly ignorable constants based on Gauss's law, as in the third order. There are some possible fourth order processes involving
only $H_{int}$ on two links, corresponding to back-and-forth hopping of fermions (see figure \ref{jumps}b), which result in constants due to the unitarity of interactions, as in the second order.

The non-constant contributions are of two types. The terms involving $H_E$ yield a renormalization to the electric Hamiltonian, of the form
\begin{equation}
\delta H_E = \frac{\mu^2 \epsilon^2}{\lambda^3} \underset{\mathbf{n},k}{\sum}\left(E_{\mathbf{n},k}\right)^2
\end{equation}
- this is the reason for the inclusion of $\chi$ fermions; Otherwise, we would have got such a renormalizations only for links which are neighbors of the "preferred" vertices of $\psi$s.

The other type contributions are the anticipated plaquette interactions. These arise from the contribution $-\mathcal{P}_0 H_{int}\mathcal{K} H_{int}\mathcal{K}H_{int}\mathcal{K} H_{int}\mathcal{P}_0$,
where the four operations of $H_{int}$ cause a fermion to hop from its "rest" vertex, around a plaquette, and back to the original place, and thus forming the desired Hamiltonian terms (see figure \ref{jumps}c).
Each plaquette operator is divided into two different orientations (clockwise and counterclockwise). Each orientation is obtained twice, because it has to start (and finish) in an even vertex, and each plaquette
contains two such vertices - i.e., each plaquette interaction is obtained once using $\psi$s and once using $\chi$s. Altogether we get the required
\begin{multline}
H_B = -\frac{2\epsilon^4}{\lambda^3} \underset{\mathbf{n}}{\sum}\left(U_{\mathbf{n},1}U_{\mathbf{n + \hat 1},2}U^{\dagger}_{\mathbf{n +  \hat 2},1}U^{\dagger}_{\mathbf{n},2} + h.c.\right) = \\
 -\frac{4\epsilon^4}{\lambda^3}\underset{\mathbf{n}}{\sum}\cos\left(\phi_{\mathbf{n},1}+\phi_{\mathbf{n + \hat 1},2}-\phi_{\mathbf{n +  \hat 2},1}-\phi_{\mathbf{n},2}\right)
\end{multline}
and eventually the abelian Kogut-Susskind Hamiltonian is obtained
\begin{equation}
H_{eff} = H_E + \delta H_E + H_B
\end{equation}

One may claim that the plaquette interactions are negligible here, since they are obtained in fourth order perturbation theory. However, note that all the terms in the effective Hamiltonian involving $\epsilon$
actually involve $\epsilon^2$ - only even orders contribute (since one can return to $\mathcal{M}_0$ only with an even number of $H_{int}$ operations). Thus, it is sufficient to demand $\epsilon^2 \ll \lambda^2$, rather than $\epsilon \ll \lambda$.

Dynamical fermions may be introduced through the inclusion of another set of fermions, with its own $H_{int}$ and $H_M$, and without any constraint (see section \ref{dynferm}).

In section \ref{2p1} we discuss a similar simulation in the real case - i.e., under real conditions, without the ideal $N_0 \rightarrow \infty$ assumption,
along with a numerical proof of principle.

\subsection{$\mathbb{Z}_N$ plaquettes}
\label{Zpl}

The effective construction of the $\mathbb{Z}_N$ Hamiltonian out of the fundamental Hamiltonian (\ref{HfundZ}) in the loop method is similar to the derivation in the cQED case. The analogy applies to the ideal limit of
the cQED simulation,
as the interactions here are exactly unitary (i.e., the link operators are unitary).
As before, one sets $\lambda$ to be the largest energy scale, comparing to $\mu$ and $\epsilon$. Initially, all the vertices are filled with bosons, as explained in the previous subsections.

In the first order we obtain $H_E$. In the second order we obtain an ignorable constant, thanks to the unitarity of the elementary interactions.

In the third order, as in the $U(1)$ case, one obtains from the anti-commutator contribution (see section \ref{Upl})
$-\frac{4 \alpha \epsilon^2}{\left(2\lambda\right)^2} \mathcal{N}H_E$ (where $\mathcal{N}$ is the number of
links, and $\alpha = -\mu/2$). From the other
contribution, one obtains for each link
\begin{equation}
\frac{2\alpha \epsilon^2}{\left(2\lambda\right)^2} \left(  2 H_E + \left(QPQ^{\dagger} + QP^{\dagger}Q^{\dagger} + Q^{\dagger}PQ + Q^{\dagger}P^{\dagger}Q\right)\right)
\end{equation}
where the $2$ comes from the bosonic creation and annihilation operators of the vertices. The $\mathcal{N}H_E$ dependent terms from both contributions cancel.
Using $QPQ^{\dagger} = e^{i \delta}P,QP^{\dagger}Q^{\dagger} = e^{-i \delta}P^{\dagger}$ ($\delta = \frac{2\pi}{N}$, see section \ref{ZN}) we get a renormalization to $H_E$:
\begin{equation}
\delta ^{\left(3\right)}H_E = \frac{ \epsilon^2 \mu}{\lambda^2}\sin^2\left(\frac{\delta}{2}\right)\underset{l}{\sum}\left(P_l + P_l^{\dagger}\right)
\end{equation}

In the fourth order, the non-constant contributions are two:
\begin{enumerate}
    \item{An "undesired" term (which is still gauge invariant, of course):
    \begin{equation}
    H'_E = \frac{\epsilon^2 \mu^2}{2 \lambda^3}\cos\left(\delta\right)\sin^2\left(\frac{\delta}{2}\right)\underset{l}{\sum}\left(P_l^2 + P_l^{\dagger 2}\right)
    \end{equation}
    This term becomes "good" $N=2,3$, since for $N=2$, $P^2 = P^{\dagger 2} = 1$ (and thus it is a constant),
    and for $N=3$, $P^2 = P^{\dagger}$ and $P^{\dagger 2} = P$ and thus it is yet another renormalization of $H_E$.
    }
    \item{The anticipated plaquette terms:
    \begin{equation}
    H_B = -\frac{4 \epsilon^4}{\lambda^3}\underset{\mathbf{n}}{\sum}\left( Q_{\mathbf{n},1} Q_{\mathbf{n + \hat 1},2} Q^{\dagger}_{\mathbf{n + \hat 2},1} Q^{\dagger}_{\mathbf{n},2} + h.c.\right)
    \end{equation}
    }
\end{enumerate}

Thus, eventually, if we define a renormalized $\mu_{ren} = \mu \left(1 - \frac{2 \epsilon^2}{\lambda^2}\sin^2\left(\frac{\delta}{2}\right)\right)$ and the appropriate $H_{E,ren} = H_E + \delta ^{\left(3\right)}H_E$, we obtain to fourth order, the desired $\mathbb{Z}_N$ lattice gauge theory Hamiltonian, with some corrections:
\begin{equation}
    H_{eff} = H_{E,ren} + H_B + H'_E
\end{equation}
where for $N \rightarrow \infty$, $H_{E,ren} \rightarrow H_E$ and $H'_E \rightarrow 0$, for $N=2$ $H'_E$ is an ignorable constant, and for $N=3$ $H'_E$ is another renormalization of into $H_{E}$, and thus in these
two cases the simulation is exact.

\subsection{$SU(N)$ plaquettes}
\label{Spl}

Finally, we shall describe the effective construction of an $SU(N)$ gauge theory, using the appropriate elementary interactions (for the ideal case, in which they really contain unitary matrices), with the loop method. Again, for simplicity, we describe the $2+1$ dimensional case. We start with the Hamiltonian
\begin{equation}
H = H_E + H_{int} + H_C
\label{HNAF}
\end{equation}
where the electric Hamiltonian of an $SU(N)$ LGT as in equation (\ref{HEgen}) (again, we define $\mu = \frac{g^2}{2}$, and $H_{int}$ is the appropriate elementary interaction (\ref{fbhop}), with some fixed representation $r$ of
the group. We choose it to be the fundamental representation, and for simplicity we shall next drop the representation index $r$ when referring to the fundamental representation.

The spinors will be in the size of the representation - thus, for the fundamental representation of $SU(N)$ we need $N$ fermionic species. Besides interacting with the links in $H_{int}$, they are also constrained by $H_C$,
which is similar to the previous constraining Hamiltonians. One such constraining Hamiltonian is
\begin{equation}
H_C = -\lambda \underset{v \text{ special}}{\sum}{\psi^{\dagger}_v \psi_v} =  -\lambda \underset{v \text{ special},a}{\sum}{\psi^{\dagger}_{v,a} \psi_{v,a}}
\end{equation}
where we define the "special" vertices to be the ones with both indices even: i.e., it is energetically favorable for the fermions to be in special vertices.
As before, we introduce another set of fermions, $\chi$, with similar $H_{int}$, $H_c$ constraining to other special vertices - with both the indices odd, and no interactions with the $\psi$s.
 Each plaquette contains exactly two such vertices, one of each type (even or odd).

Initially, we prepare the system in one of two possible classes - either a pure state of the form
\begin{equation}
\left|sys\right\rangle = \left|\Psi\right\rangle \otimes \left|\{U\}\right\rangle
\end{equation}
where $\left|\{U\}\right\rangle$ is a bosonic state and $\left|\Psi\right\rangle$ is the state of the fermions, in which the the "non-special" vertices contain no fermions at all, while the
 "special" ones are fully occupied,
or the mixed state
\begin{equation}
\rho_{sys} = \rho_{\Psi} \otimes \left|\{U\}\right\rangle \left\langle\{U\}\right|
\end{equation}
where $\left|\{U\}\right\rangle \left\langle\{U\}\right|$ is the density matrix of some pure state of the bosons, and $\rho_{\Psi}$ is a mixed state, in which the "non-special" vertices contain no fermions at all, while the
 "special" ones are each prepared in
\begin{equation}
\rho = \frac{1}{N}\underset{a}{\sum} \left|a\right\rangle \left\langle a\right|
\end{equation}
(where $\left|a\right\rangle$ corresponds to a state of a single $a$ fermion, $\psi$ or $\chi$, depending on the site).
In both the possibilities the initial state should be gauge invariant, of course.

As before, we set $\lambda$ to be the largest energy scale in the Hamiltonian, and construct an effective Hamiltonian for its ground sector $\mathcal{M}_0$.
In the case where the fermions are prepared in a mixed state, their
"tracing out" will literally be tracing out, i.e. the effective Hamiltonian will be the result of a partial trace over the fermionic degrees of freedom,  of the perturbative expansion,
$\text{Tr}_{\text{F}}\left(H_{eff} \rho_{\Psi}\right)$

\subsubsection{Effective non-plaquette terms}
In the first order, we obtain $H_E$. Then we are left with an effective Hamiltonian, acting on the pure state of bosons.
In the second order, we get once again an ignorable constant, due the the unitarity of the interactions.

We give here the derivation of terms of the initial mixed-state case, however, the final results of the initial pure-state case are very similar. Thus, we shall introduce for the final results the symbol $\xi_{in}$, which equals $1$ if the initial state is pure, and $1/N$ if it is mixed.

\begin{widetext}
The third order terms are $H_{eff}^{\left(3\right)} = \mathcal{P}_0 H_1\mathcal{K} H_1 \mathcal{K} H_1\mathcal{P}_0 - \frac{1}{2}\left\{\mathcal{P}_0 H_1\mathcal{K}^2 H_1\mathcal{P}_0,\mathcal{P}_0 H_1\mathcal{P}_0\right\}$.
Let us examine the first part. Consider a special vertex $1$, some link emanating from it (in the following example - negative) and the vertex on its other edge $2$. We act with the terms constructed out of these components
on a an element of a state in $\mathcal{M}_0$:
\begin{equation}
\frac{\epsilon^2}{\lambda^2}\underset{a,b,c,d,e,f}{\sum}\left(\psi_1^{\dagger}\right)_a \left(U^{\dagger}\right)_{ab} \left(\psi_2\right)_b \left|0_1 e_2\right\rangle \left\langle 0_1 e_2\right| H_E
\left|0_1 f_2\right\rangle \left\langle 0_1 f_2\right| \left(\psi_2^{\dagger}\right)_c U_{cd} \left(\psi_1\right)_d \left|d_1 0_2\right\rangle \left\langle d_1 0_2\right|
\end{equation}

$H_E$ does not involve any fermions, and thus $\left\langle 0_1 e_2\right| H_E \left|0_1 f_2\right\rangle = \delta_{ef}H_E$; From the fermionic terms we get
\begin{equation}
\left(\psi_1^{\dagger}\right)_a  \left(\psi_2\right)_b \left|0_1 e_2\right\rangle \left\langle 0_1 e_2\right| \left(\psi_2^{\dagger}\right)_c \left(\psi_1\right)_d \left|d_1 0_2\right\rangle \left\langle d_1 0_2\right| = \delta_{be}\delta_{ec} \left|a_1 0_2\right\rangle \left\langle d_1 0_2\right|
\end{equation}
Considering the entire local fermionic ensemble, and tracing it out, we get $\frac{1}{N} \delta_{bc}\delta_{ad}$, and eventually the contribution for each negative link $\frac{\epsilon^2}{N\lambda^2} U^{\dagger}_{ab}H_E U_{ba}$,
and similarly, for a positive one - $\frac{\epsilon^2}{N\lambda^2} \underset{ab}{\sum} U_{ab}H_E U^{\dagger}_{ba}$.

\begin{equation}
U^{\dagger}_{ab}H_E U_{ba} = U^{\dagger}_{ab}\left(H_E - \mu \mathbf{E}^2 \right)U_{ba} + \mu U^{\dagger}_{ab} \mathbf{E}^2 U_{ba} = \delta_{aa}\left(H_E - \mu \mathbf{E}^2 \right) + \mu U^{\dagger}_{ab} \mathbf{E}^2 U_{ba}
\end{equation}
where $\mathbf{E}^2$ is the Casimir operator of the relevant link. Using the commutation relations of group elements with group generators (see section \ref{SUN}), we get
\begin{equation}
\underset{ab}{\sum}\left(\delta_{aa}\left(H_E - \mu \mathbf{E}^2 \right) + \mu U^{\dagger}_{ab} \mathbf{E}^2 U_{ba}\right) =
 \mu \text{Tr}\left(\mathbf{\lambda}^2\right) + N H_E + \mu \underset{a}{\sum}\left\{E_a,\text{Tr}\left(T_a\right)\right\}
\end{equation}
where $T_a$ is the matrix representation of $E_a$ within the chosen representation, which for $SU(N)$ is traceless.
\end{widetext}

For positive links, the only difference will be a change of sign to the last term, which is zero, so their contribution is the same. Altogether we get, neglecting the constant terms of $\text{Tr}\left(\mathbf{\lambda}^2\right)$,
$4\mathcal{N}_V \frac{\epsilon^2}{\lambda^2} H_E$, where $\mathcal{N}_V$ is the number of "special" even vertices (the $4$ factor is due to the fact, that in two spatial dimensions, there are 4 links neighboring each vertex).

The other third order contribution is of the form $-4\mathcal{N}_V \frac{\epsilon^2}{\lambda^2} H_E$, and thus the entire third order contribution (neglecting ignorable constants) is zero.
This is all thanks to the tracelessness of the matrix representations of $SU(N)$'s generators. Suppose we extended our gauge group to $U(N)$ - then, we would have generators with nonzero trace, which would lead to the Gauss's
law terms and renormalizations, as in the $U(1)$ case. The $\chi$ fermions will give rise, of course, to a similar contribution.

The fourth order terms involving double operation of $H_E$, will similarly result in constants and cancelled terms, thanks to the tracelessness of $SU(N)$ generators. However, here we have one nonzero contribution, which is
a renormalization to the electric part (for both the initial states possibilities)
\begin{equation}
\delta H_E = -\frac{4 \xi_{in} \mu^2 \epsilon^2}{\lambda^3}C\left(r\right)\underset{\mathbf{n},k,a}{\sum} \left(E_{\mathbf{n},k}\right)_a \left(E_{\mathbf{n},k}\right)_a
\end{equation}
where for the fundamental representation of $SU(N)$, $C\left(N\right) = 1/2$.

Fourth order contributions which go back and forth with $H_{int}$ on two links give rise to constants as well,
due to the unitarity.

\subsubsection{The plaquette terms}
Finally, we shall construct the plaquette interactions, which come, as in the $U(1)$ case, from transferring auxiliary fermions around a plaquette. Here, after eliminating the fermionic degrees of freedom in the non-special vertices,
 one effectively gets, for example, terms like
\begin{equation}
-\frac{\epsilon^4}{\lambda^3}\underset{abcde}{\sum}\left(\psi_v^{\dagger}\right)_a \left(U_1\right)_{ab} \left(U_2\right)_{bc} \left(U_3^{\dagger}\right)_{cd} \left(U_4^{\dagger}\right)_{de} \left(\psi_v\right)_{e}\left|e\right\rangle \left\langle e \right|
\end{equation}
(with the conventions of figure \ref{plaqint}, with $v$ the vertex labeled there by $\mathbf{n}$).
By tracing out the fermions, we get $\delta_{ae}$, which "closes" the plaquette and introduces the desired group-trace. This yields the plaquette Hamiltonian
\begin{equation}
H_B = -\frac{2 \xi_{in} \epsilon^4}{\lambda^3} \underset{\text{plaquettes}}{\sum} \left(\text{Tr}\left(U_1 U_2 U^\dagger_3 U^\dagger_4\right) + h.c. \right)
\end{equation}
(the factor 2 is due to the two types of auxiliary fermions) and eventually we get effectively, up to fourth order, the $SU(N)$ pure-gauge Kogut-Susskind Hamiltonian,
\begin{equation}
H_{eff} = H_E + \delta H_E  + H_B
\end{equation}
without any corrections ($\delta H_E$ is just a renormalization)! Moreover, the leading order of $\epsilon$ is the fourth one, and thus here it is even sufficient to demand $\epsilon^4 \ll \lambda^4$.

One should also note, that by introducing other set of fermions, interacting with the bosons with a similar $H_{int}$ but non-constrained, one can introduce dynamic fermions to the system. This is discussed in section \ref{dynferm}.

\section{Quantum simulation of $2+1$-d gauge theories}
\label{2p1}

Putting together the elementary link interactions of section \ref{elemint}, and the loop method for the plaquette interactions of section \ref{plaqs}, one could construct quantum simulations of lattice gauge theories in $2+1$ dimensions. However, in several cases one would have to face "real" conditions, instead of the "ideal" conditions of the previous section, such of the use of finite Hilbert spaces instead of infinite ones.

In that sense, the quantum simulation of $\mathbb{Z}_N$ is the most accurate: in this theory, the local hilbert spaces are already of finite dimension, and thus nothing should be added to the discussion of these models in the previous section.

On the other hand, quantum simulation of the continuous gauge theories - $U(1)$ and $SU(N)$ - require some truncation of the Hilbert space (This was already mentioned in the case of the Schwinger model, in section \ref{Sch}).
In the case of cQED, the simulation requires the use of BECs. In this section we show that it should still work with a
finite number of bosons per link, derive the conditions for that, and give numerical evidence. Finally we shall comment on the $SU(2)$ $2+1$-d simulation.

\subsection{$2+1$-d simulation of cQED using a finite number of bosons}
\label{finbos}
\begin{figure*}[t!]
  \centering
  \includegraphics[scale=1]{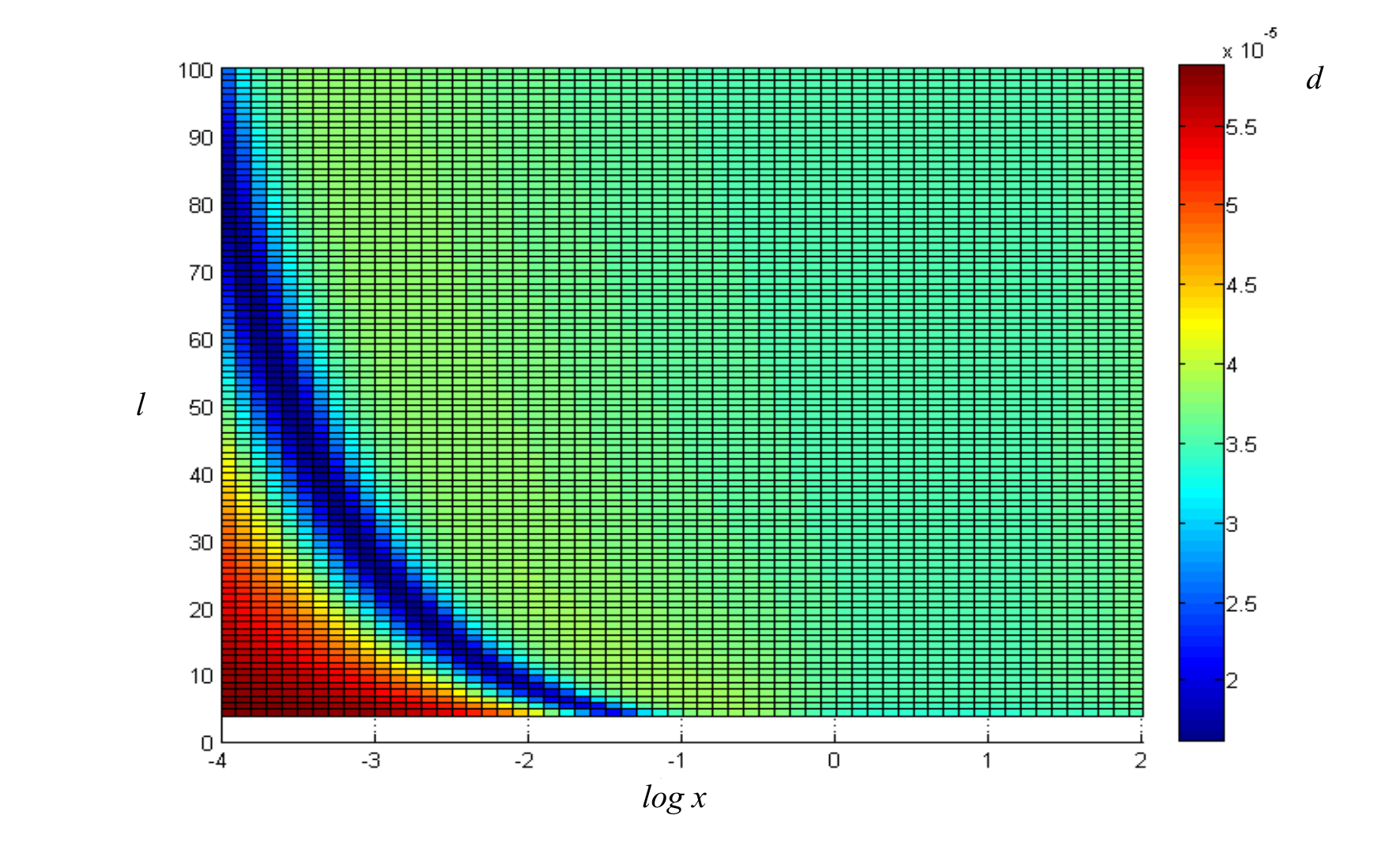}
  \caption{A plot of $d$ (equation (\ref{stdev})) for several values of $\ell$,$x$. One can see that $d$ grows as $\ell$,$x$ decrease. However, in this very same region of parameters, the credibility of it is low, as explained
  in the text.}
  \label{stdplot}
\end{figure*}

\begin{figure*}[t!]
  \centering
  \includegraphics[scale=0.8]{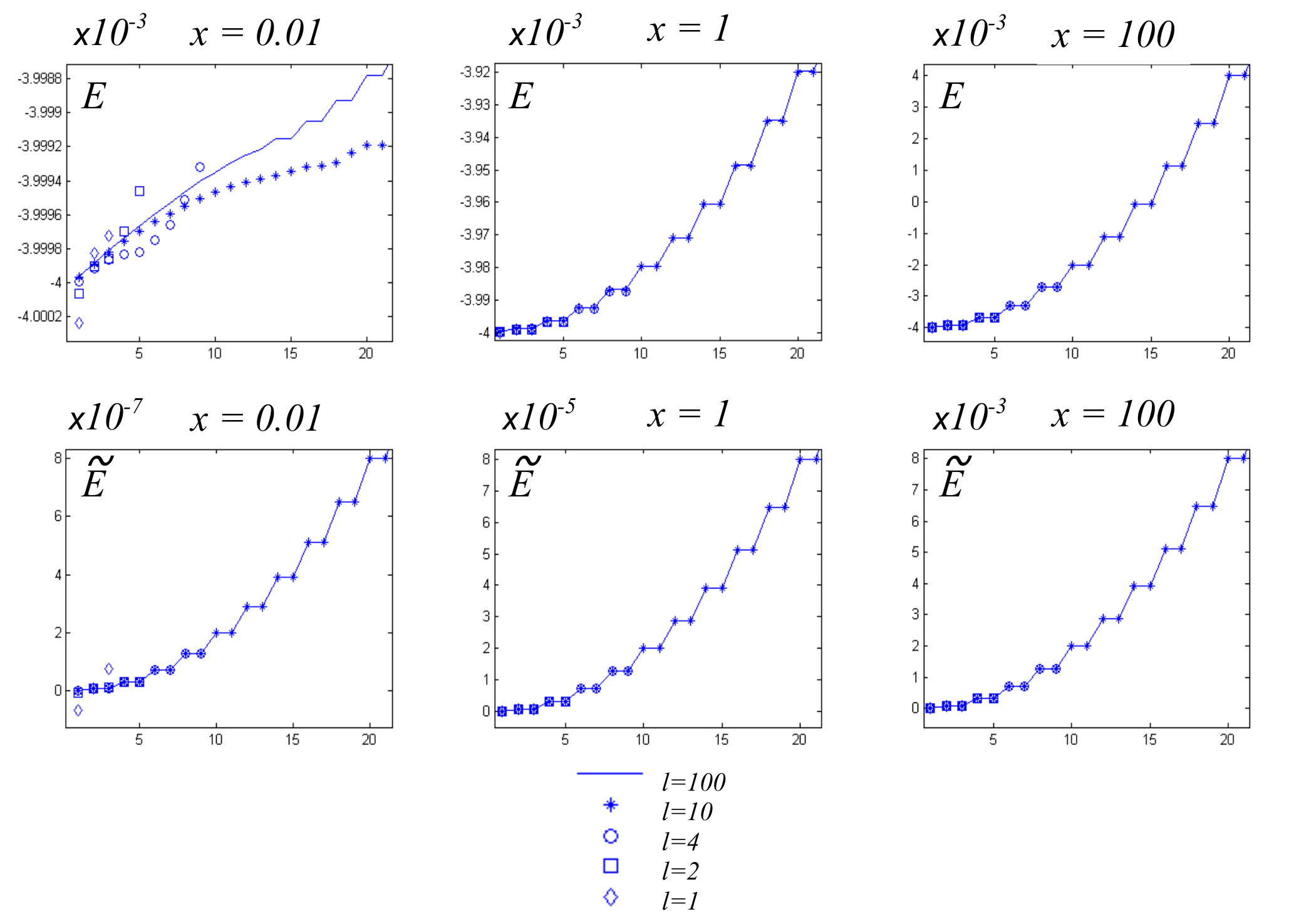}
  \caption{Plots of the spectra of $H$ (upper panel) and $\tilde H$ (lower panel) for small values of $\ell$ and several values of $x$, compared to the fairly reasonable approximation of the non-truncated Kogut-Susskind model
  using $\ell=100$. Note that the qualitative correspondence of the two spectra arises as $x, \ell$ increase.}
  \label{spectra}
\end{figure*}

\begin{figure*}[t!]
  \centering
  \includegraphics[scale=1]{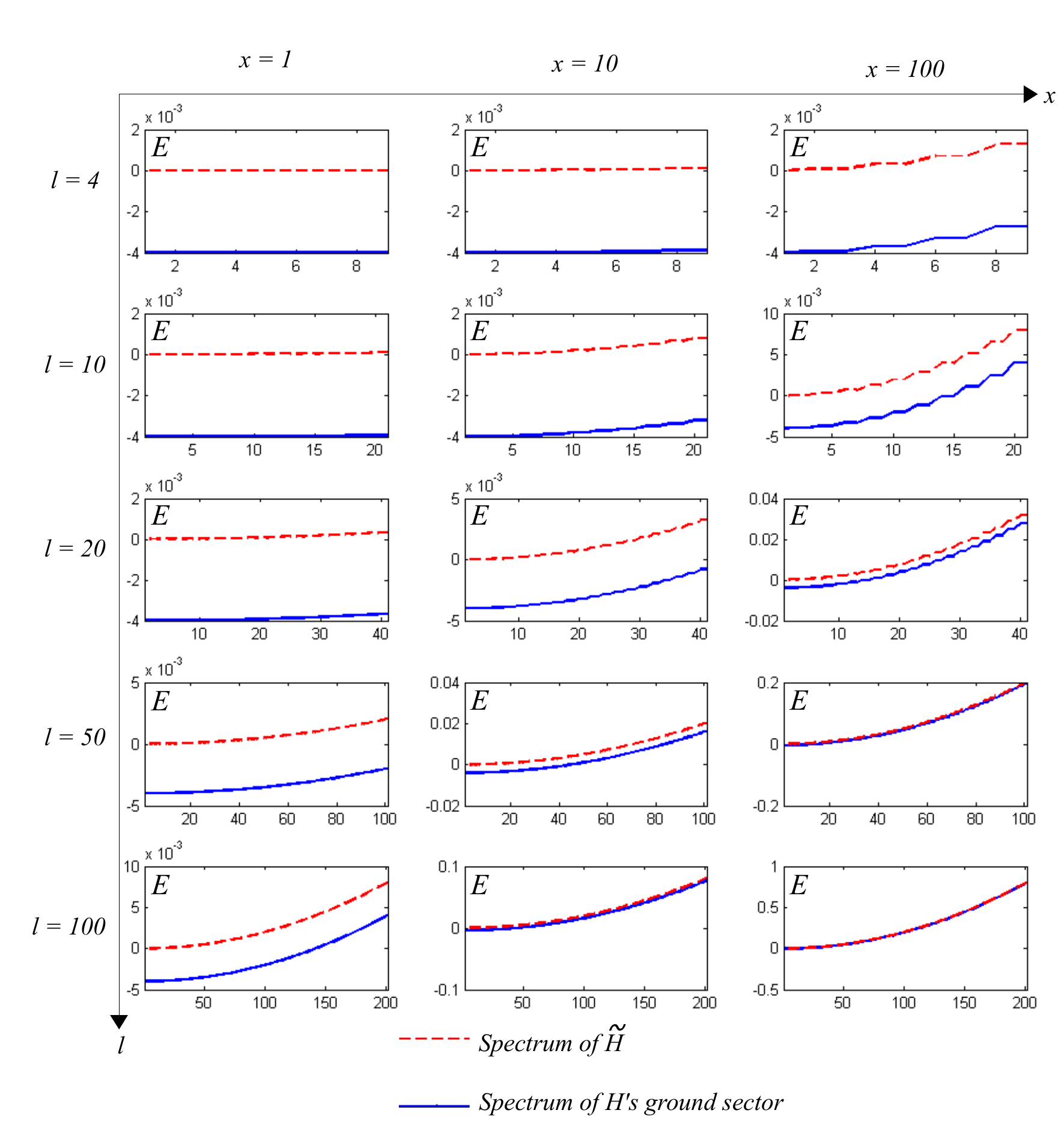}
  \caption{Plots of the spectra of $H, \tilde H$, enabling a qualitative impression on the constant difference between them, for several valus of the simulation parameters $x$,$\ell$.}
  \label{goodgraphs}
\end{figure*}

We have shown in section \ref{Upl} how to obtain, effectively, the plaquette interactions using the already gauge invariant elementary interactions. However, in the derivation we have used unitary matrices in the elementary interactions ($N_0 \rightarrow \infty$),
whereas in the "real" scenario $N_0 = 2\ell$ is finite, even if it's large, and thus the interactions contain angular momentum ladder operators (\ref{HintS}) which are nonunitary. Here we shall describe how to handle
these effects in order to achieve an accurate simulation despite the non-unitarity.

We still include the $\chi$ fermions, with the same constraining part (\ref{Hc}), but now $H_{int}$ takes the form
\begin{equation}
H_{int} = \frac{\epsilon}{\sqrt{\ell \left(\ell + 1 \right)}} \underset{\mathbf{n},k}{\sum}\left(\psi_{\mathbf{n}}^{\dagger} L_{+,\mathbf{n},k}  \psi_{\mathbf{n+\hat{k}}} + \chi_{\mathbf{n}}^{\dagger} L_{+,\mathbf{n},k}  \chi_{\mathbf{n+\hat{k}}} + h.c.\right)
\end{equation}

\subsubsection{First and second order contributions}
\label{real1first}
In the first order, we get $H_E$ as before.

In the second order, we get the same type of contributions, but now they will not be constant anymore. What we have now are contributions of the form
\begin{equation}
-\frac{\epsilon^2}{\lambda}\frac{L_{\pm}L_{\mp}}{\ell \left(\ell + 1 \right)} = -\frac{\epsilon^2}{\lambda}\left(1 - \frac{L_z^2}{\ell \left(\ell + 1 \right)} \pm \frac{L_z}{\ell \left(\ell + 1 \right)}\right)
\end{equation}
(where $L_+L_-$ is for positive links, and vice-versa for the negative ones). This contribution becomes constant as $\ell \rightarrow \infty$, because then, any eigenvalue of $L_z$
satisfies $m \ll \ell$. In the case of a finite $\ell$, one can neglect the first constant, and after summing on all the links, obtain a renormalization factor for the electric
Hamiltonian from the $L_z^2$ term, and a Gauss's law (which is an ignorable constant) from the linear $L_z$ part which has the correct signs. Thus, second order leads to a
renormalization of $H_E$ (and this is why the $\chi$s are important).

\subsubsection{Third and fourth order contributions}
Here, after cancelling the equal part in the two possible contributions (see the previous section), we are left with
\begin{equation}
\frac{\mu \epsilon^2}{\lambda^2} \left(1 \mp 2L_z\right)\frac{L_{\pm}L_{\mp}}{\ell \left(\ell + 1 \right)}
\end{equation}
where the choices of signs for positive/negative links are as in the second order.

Again, in the ideal limit, $\frac{L_{\pm}L_{\mp}}{\ell \left(\ell + 1 \right)} \underset{\ell \rightarrow \infty}{\longrightarrow} 1$, and then we get a constant + Gauss's laws (which are constants too).
However, for a finite $\ell$, the contributions are again non constant: besides the constants, we get linear terms in $L_z$, which correspond to Gauss's law, and a third order renormalization of $H_E$,
but also $L_z^3$ terms.
If we work with $\ell =N_0/2= 1$, $L_z^3=L_z$ and thanks to the correct signs get Gauss's law again. However, this is not a very interesting and obviously not the general case, and thus we wish to find some
way to deal with these extra terms. We shall first focus on the fourth order contributions, and then conclude what to do about the "problematic" terms from both orders.

In the fourth order there are three types of contributions.
First, the plaquette terms, which now take the form
\begin{multline}
H_B = -\frac{2\epsilon^4}{\lambda^3 \ell^2 \left(\ell+1\right)^2} \times \\ \underset{\mathbf{n}}{\sum}\left(L_{+,\mathbf{n},1}L_{+,\mathbf{n + \hat 1},2} L_{-,\mathbf{n + \hat 2},1}L_{-\mathbf{n},2} + h.c.\right)
\end{multline}

The second type of terms involve operations with $H_{int}$ only, and they include, after the reduction of constants, products of $L_z$ and $L_z^2$ of neighboring links intersecting in odd vertices. This
term is in the same order of the plaquette terms - $O\left(\frac{\epsilon^4}{\lambda^3} \right)$. These terms, which we call $H'_B$, are, indeed, unwanted terms, but can still be tolerated, as they are
gauge invariant and not stronger than the plaquette interactions.

The third terms take the form $-\frac{\epsilon^2\mu^2}{\lambda^3 \ell \left(\ell+1\right)}\left\{L_z,L_{\pm}\right\}\left\{L_z,L_{\mp}\right\}$ on positive/negative links. These terms will renormalize
 $H_E$
in the ideal limit, but for a finite $\ell$ will include, besides the constants, $L_z$ and $L_z^2$ terms, which are "treatable", also $L_z^3$ and $L_z^4$ terms (again, with the correct signs to
contribute to
Gauss's law and renormalize $H_E$ for $\ell=1$). In order to eliminate the effect of these terms, along with the undesired third order terms, we define $\mu \equiv \beta \frac{\epsilon^2}{\lambda}$,
where $\beta \leq 1$ is a
dimensionless parameter. Then, the undesired third order terms are $O\left(\frac{\epsilon^4}{\lambda^3} \right)$ - like the plaquettes, and thus can be tolerated, at least for states which are
superpositions of $m \ll \ell$
mostly. The fourth order undesired terms become effectively sixth order terms - $O\left(\frac{\epsilon^6}{\lambda^5} \right)$ - and thus can be safely neglected.

Thus, for a finite $N_0 = 2\ell$, if the parameters are tuned correctly, one gets effectively, up to constant, the Hamiltonian
\begin{equation}
H_{\ell} = \tilde H_E + H_B + O\left(\frac{\epsilon^4}{\lambda^3} \right)
\label{Hl}
\end{equation}
with
\begin{equation}
\tilde H_E = \left( \frac{\beta}{\lambda} + \frac{1}{\ell\left(\ell + 1 \right)} \right) \frac{\epsilon^2}{\lambda} \underset{\mathbf{n},k}{\sum}\left(E_{\mathbf{n},k}\right)^2
\end{equation}
which is expected to give rise to the same dynamics as the Kogut-Susskind Hamiltonian $H = \tilde H_E + H_B$, at least for a regime in which $\frac{\epsilon^4}{\lambda^3}$ is small enough: i.e.,
not in the extreme
weak limit, but apparently not only in the strong limit, but also in a regime where $H_E,H_B$ are of the same order of magnitude. The two Hamiltonians have to differ by a constant, at least for
states which are
superpositions mostly of $m \ll \ell$.

In order to see that, we shall consider some numerical results. But before that, let us point out two important issues about the Hamiltonian parameters. Although the relation
$\mu \equiv \beta \frac{\epsilon^2}{\lambda}$
might seem to imply that $H_E$ will always be stronger than $H_B$, because of the powers of $\epsilon$ and $\lambda$, one should note that thanks to $\beta$, which can be chosen small, one
can still go to weaker regimes.
Moreover, as in the ideal case, $H_{int}$ only contributes to the effective Hamiltonian in even repetitions, and thus, again, it is sufficient to demand $\epsilon^2 \ll \lambda^2$ rather than
$\epsilon \ll \lambda$.

\subsection{$2+1$-d cQED simulation - A Numerical proof of principle}
\label{numsim}

In order to see whether the "real" case simulation still yields valuable results, one has to check the effects of the $O\left(\frac{\epsilon^4}{\lambda^3}\right)$ in $H_{\ell}$ (\ref{Hl}).
This can be done if one compares the spectrum of the "desired" Hamiltonian,
\begin{equation}
\tilde H_{\ell} = \tilde H_E + H_B
\label{Hdes}
\end{equation}
with the spectrum of the lower energy sector (i.e., the states fulfilling the constraint) of the fundamental Hamiltonian,
\begin{equation}
H = H_E + H_{int} + H_C
\end{equation}

The accuracy of the simulation will be deduced from the observation of a constant energy shift between the spectra of the two Hamiltonians. This will mean that both of them give rise to the same
dynamics.
Such results will trivially take place in the strong coupling limit, where $\epsilon \rightarrow 0$. However we wish to check what happens in other coupling regimes.

We have run numerical simulations of a single plaquette, using the parameters $\lambda = 10, \epsilon = 0.1$. We changed the values of $\ell$, $\beta$, using the convenient dimensionless ratio between
the scales of $H_E$ and $H_B$,
\begin{equation}
x = \frac{1}{2}\left(\frac{\beta}{\lambda} + \frac{1}{\ell\left(\ell+1\right)}\right)\frac{\lambda ^2}{\epsilon^2}
\end{equation}
parametrizing the coupling strength. We have chosen the subspace with no (real) static charges.

We have calculated the spectra $\left\{E\right\}$ of $H$'s ground sector, as well as $\left\{\tilde E\right\}$ of $\tilde H$ for several values of $\ell$ and $x$ and calculated the value of
\begin{equation}
d \equiv \left|\frac{\text{std}\left( E - \tilde E \right)}{\text{mean}\left( E - \tilde E \right)}\right|
\label{stdev}
\end{equation}
as one can see in figure \ref{stdplot}, this value is fairly small, getting larger, as anticipated, for smaller values of the simulation parameters. However, on the other hand, if one wishes to evaluate the \emph{quality}
of simulations, it should be noted that in order to get an accurate simulation (compared to the original Kogut-Susskind model) in the weak coupling limit (small $x$) one has to pick a sufficiently large value of
$\ell$ \cite{Zohar2012}. If only the ground state is of interest, lower $\ell$s are possible. In order to achieve a better accuracy for excited states, one must increase $\ell$. This can be seen in figure \ref{spectra}.

Furthermore, as $x$ gets smaller, the energy scales of the two Hamiltonians $H, \tilde H$ separate (as can be seen in figure \ref{spectra}). Thus, the difference between them can still be treated approximately as a constant, due
to the difference of scales, but yet one should note this qualitative subtlety as well, treating the left region of figure \ref{stdplot} more carefully. On the other hand, for $x \gtrsim 1$, using sufficiently large $\ell$s, which
tend to get smaller as $x$ is increased (see figure \ref{goodgraphs}), the scales of two Hamiltonians correspond and the $d$ value is credible. As one can conclude from that, a simulation which is accurate both comparing to the
Kogut-Susskind model and with fairly constant effective Hamiltonian contributions is possible for $x \gtrsim 1$.

\subsection{$2+1$-d $SU(N)$ Yang Mills simulation}
\label{SU2d}

In the case of non-abelian theories, such as $SU(2)$, relying on \cite{NA} for the elementary interactions (as described in section \ref{Sel}),
one could simulate a version of $SU(2)$ gauge theory in which the bosonic operators in the elementary interactions are not unitary (\ref{Hinteff}).

In the $1+1$-d case we could establish a good approximation for the unitary interactions, since the elementary interactions only change $N$ by $\pm 1$, or $j$ by $\pm 1/2$ on single links. However, here, since there are plaquette interactions, we have such processes on four links at once. One can also introduce dynamic fermions, and altogether we lose the fifth-order accuracy we had in the $1+1$-d case.
Moreover, the effective Hamiltonian series may include, in this case, more terms (as we had in the "real" cQED case comparing to the ideal unitary case). This will depend on these $N$s as well.
On the other hand, note that all these "corrections" are nevertheless gauge invariant - gauge invariance is not ruined!

Therefore we conclude that due to the $\sqrt{N+1}$ operators, which would lead, already in first order in $H_{int}^{eff}$ (\ref{Hinteff}), to terms with different amplitudes than in regular $SU(2)$ theory, the \cite{NA} realization of $SU(2)$ elementary interactions allows only for simulation of $SU(2)$ gauge theories near the strong coupling limit. However, in this regime, it should reproduce the same physics, qualitatively.

\section{Inclusion of Dynamical fermions}
\label{dynferm}

We have presented methods to construct $1+1$ dimensional simulations of continuous gauge theories ($U(1)$, $SU(N)$) with dynamic matter, and have shown how, by extending to more spatial dimensions and replacing the mass Hamiltonian $H_M$ with the appropriate constraint, $H_C$, one can use the fermions as \emph{auxilliary} particles to construct effective plaquette interactions.

Here we shall discuss the ability to introduce dynamical fermions for these theories as well.
All one has to do, is to include on top of the auxilliary fermions $\psi$ and $\chi$, more fermionic species $\Psi$. These occupy the vertices as well, and the number of species should be chosen according to the group - for example, one per vertex for $U(1)$, $N$ per vertex for $SU(N)$.

The fermionic dynamics are described by the "Dirac Hamiltonian" $H_D$, which is of course gauge invariant and consists of local mass terms ($H_M$, see equation \ref{Hmass}) and elementary interactions in the form of $H_{int}$ (\ref{fbhop}). Of course, the system must be prepared such that there will not be any direct $\Psi - \psi$ and $\Psi - \chi$ interactions.

For a $U(1)$ theory in $2+1$ dimensions, for example, one should pick
\begin{equation}
\begin{aligned}
H_D = & M\underset{\mathbf{n}}{\sum}(-1)^{n_1+n_2} \Psi^{\dagger}_{\mathbf{n}} \Psi_{\mathbf{n}}  \\
+ & \gamma\underset{\mathbf{n},k}{\sum}\left(\Psi_{\mathbf{n}}^{\dagger} e^{i \phi_{\mathbf{n},k}}  \Psi_{\mathbf{n+\hat{k}}} + \Psi_{\mathbf{n+\hat{k}}}^{\dagger} e^{-i \phi_{\mathbf{n},k}}  \Psi_{\mathbf{n}} \right)
\end{aligned}
\end{equation}

The "dynamic" fermions ($\Psi$) are not constrained and thus $H_D$ appears in the first order of the effective Hamiltonian, and makes no contribution to the second one.

In the third and fourth order one has to be more cautious. First, suppose that the elementary interactions contain unitary operators (as in section \ref{plaqs}). There,
\begin{equation}
\left[H_D,H_{int}\right]=0
\end{equation}
and thus $H_D$ makes no contribution to the third and fourth orders (as well as to higher ones) and we get the desired results.

In case the interactions are not exactly unitary (as in the models of section \ref{2p1}), there will be some nonvanishing contributions in higher orders. However, they will all be gauge invariant, and introduce only small corrections to the desired interactions.
For example, in a $U(1)$ (cQED) simulation with finite number of bosons (as in section \ref{finbos}), where $H_D$ takes the form
\begin{equation}
\begin{aligned}
H_D =  & M\underset{\mathbf{n}}{\sum}(-1)^{n_1+n_2} \Psi^{\dagger}_{\mathbf{n}} \Psi_{\mathbf{n}} \\
& + \frac{\gamma}
{\sqrt{\ell \left(\ell + 1\right)}}
\underset{\mathbf{n},k}{\sum}\left(\Psi_{\mathbf{n}}^{\dagger} L_{+,\mathbf{n},k}  \Psi_{\mathbf{n+\hat{k}}} + \Psi_{\mathbf{n+\hat{k}}}^{\dagger} L_{-,\mathbf{n},k}  \Psi_{\mathbf{n}} \right)
\end{aligned}
\end{equation}
one obtains third order corrections of the form
\begin{equation}
\frac{\epsilon ^2 \gamma}{\lambda^2} \left( \Psi_{\mathbf{n}}^{\dagger} L_{+,\mathbf{n},k} \frac{L_{z,\mathbf{n},k}}{\left(\ell\left(\ell + 1\right)\right)^{3/2}} \Psi_{\mathbf{n+\hat{k}}} + h.c. \right)
\end{equation}
for a positive link, and
\begin{equation}
-\frac{\epsilon ^2 \gamma}{\lambda^2} \left( \Psi_{\mathbf{n}}^{\dagger} \frac{L_{z,\mathbf{n},k}}{\left(\ell\left(\ell + 1\right)\right)^{3/2}} L_{+,\mathbf{n},k}  \Psi_{\mathbf{n+\hat{k}}} + h.c. \right)
\end{equation}
for a negative one.
This is a negligible (third order) correction to $H_D$ - but is gauge invariant, of course.

Another consequence of the introducing dynamic charges is that the divergence of electric field is no longer a constant of motion - it is now equal to the dynamic charge on the vertex, which is dynamic. The effective terms in the $U(1)$ simulation include such divergences, which have been previously, for static charges, dismissed as constants in the Hamiltonian. Now this is not the case, and these terms should be taken into account. However, we only obtain this "Gauss's Law" contributions at "special vertices" (where the auxilliary fermions are constrained to be). This introduces an asymmetry to the system. Nevertheless, Gauss's law is still satisfied (as the symmetry is not broken), and thus the spectrum is still divided into sectors of \emph{static} charges. If one initially prepares the system in a state where there are no static charges, only the dynamical ones contribute to Gauss's law and thus these local divergences of the electric field are proportional to the charge, which is expressed in terms of the local $\Psi$ number in the "special vertices". Thus one can introduce counter terms, proportional to
$\Psi^{\dagger} \Psi$ on special vertices to eliminate this asymmetry. Note that these are merely corrections to the mass $M$ in $H_M$ in the special vertices.

For a completely unitary theory, these Gauss's law terms start to appear only in the third order (see \ref{unit1third}) and thus the required counter terms are initially very weak and thus their contribution to the effective series may be taken into account only in the first order (where it is required). In a "real" theory with angular momentum operators, such "Gauss's laws" appear already in the second order (see \ref{real1first}), the counter terms have to be a little stronger and in order to have the possibility to disregard their contributions to elements in the effective series in orders higher than 1, the interactions parameters must be chosen carefully.

\section{Summary and Conclusions}

In this paper we have presented a new method for the simulation of lattice gauge theories in high energy physics, using ultracold atoms in optical lattices.
Unlike previous proposals, where the gauge invariance was obtained effectively, here the gauge invariance is fundamentally in the atomic Hamiltonian, without using perturbation theory.
This is done by utlizing the conservation of hyperfine angular momentum, introducing the gauge invariant elementary matter-gauge field interactions along links.

At a second stage, the nonlinear self-interactions of the gauge fields along plauqettes are constructed effectively, using the introduced loop method, from the already gauge invariant Hamiltonian with the elementary interactions. Although the derivation uses
fourth order perturbation theory, it is effectively to second order, due to the vanishing of the relevant odd orders.

Along the simulation proposals in these paper, was the suggestion to simulate compact QED in $1+1$ dimensions - the Schwinger model. This simulation scheme utilizes in this case a small number of atomic species and does not involve perturbative
and effective methods at all. Thus it could serve as the first step towards the realization of quantum simulations of gauge theories and high energy physics models.

\section*{Acknowledgements}
EZ is supported by the Adams Fellowship Program of the Israel Academy of Sciences and Humanities.
IC is partially supported by the EU project AQUTE.
BR acknowledges the support of the Israel Science Foundation, the German-Israeli Foundation, and the European Commission (PICC).

\section*{Appendix: Further details on the lattice gauge theories discussed in the paper}

\subsection*{1. Confinenemt in cQED}

In this appendix we briefly review confinement in $cQED$, whose Hamiltonian was discussed in section \ref{cQED}.

Let us consider first the states of the pure gauge theory, i.e. with no dynamical matter. In this case only static charges are possible, and the Hamiltonian is the abelian Kogut-Susskind Hamiltonian (\ref{aHKS}). Suppose we use the
local flux basis - i.e., the basis consisting of products of flux states over the links of the entire lattice. In the \emph{strong coupling limit}, where $g^2 \rightarrow \infty$, this basis is a good choice, as its elements are
eigenstates of $H_E$. Note, however, that these eigenstates are divided into several sectors, depending on the static charges - eigenvalues of $G_{\mathbf{n}}$, as these are constants of motion, setting the Gauss's law (\ref{aGauss}).
 Let us discuss this limit, taking $H_B$ as a perturbation to $H_E$.

Consider, in the strong limit, the case of a single charge $Q_{\mathbf{n}}=1$ in some vertex $\mathbf{n}$. In order to respect Gauss's law, an infinite string
of flux must be introduced to the system. In fact, the (degenerate) zeroth order ground state of this charge sector is already of infinite energy: an infinite string of flux $\pm 1$ must emanate from $\mathbf{n}$. This is an evidence
 of charge confinement in the strong limit: a single charge "costs" infinite energy.
Thus we must consider bound states of at least two charges. Let us have a look on the state of two such charges, say $Q_{\mathbf{n}}=1$ and $Q_{\mathbf{n}+R\mathbf{\hat k}}=-1$. There, the zeroth order eigenstate consists of a string
 of flux $1$, connecting the charges, i.e. the state
\begin{equation}
\left|R^{\left(0\right)}\right\rangle = \psi^{\dagger}_{\mathbf{n}}\underset{i=0}{\overset{R-1}{\prod}}e^{i \phi_{\mathbf{n}+i\mathbf{\hat k}}}\psi_{\mathbf{n}+R\mathbf{\hat k}}\left|vac\right\rangle
\label{fluxtube}
\end{equation}
where $\psi^{\dagger}_{\mathbf{n}}$, $\psi^{\dagger}_{\mathbf{n}}$ create positive/negative static charges in the appropriate vertices and $\left|vac\right\rangle$ is the zeroth order chargeless ground state, with flux zero everywhere.
 Note that this state is gauge invariant (just check the abelian version of the local gauge transformation laws (\ref{psitrans},\ref{utrans}). This state describes an electric flux tube, as expected within a confining phase.
 The energy of such a state, to zeroth order, is just the eigenvalue of $H_E$:
\begin{equation}
H_E \left|R^{\left(0\right)}\right\rangle = \frac{g^2}{2} R \left|R^{\left(0\right)}\right\rangle
\end{equation}
- the static energy of this particle-antiparticle bound pair (a "meson") is proportional to $R$ (up to higher order corrections, $O\left(g^{-4}\right)$. This manifests the linear potential law, expected in a confining phase:
$V\left(R\right) \propto R$, unlike the well-known Coulomb phase behavior, $V\left(R\right) \propto R^{-1}$.

The weak limit $g^2 \rightarrow 0$, however, is different. There, $H_B$ is stronger, and these flux states are no longer eigenstates of the system, even not approximate ones. In $3+1$ dimensions, there is a phase transition to a
 Coulomb phase \cite{KogutLattice}. In $2+1$ dimensions, the theory confines for all values of the coupling constant, and it is a non-perturbative effect \cite{Polyakov,BanksMyersonKogut,DrellQuinnSvetitskyWeinstein,BenMenahem}.
  This is also the case for $1+1$ dimensions, where the theory is exactly solvable. When temperature enters the game as well, there is a phase transition to a Coulomb phase for small $g$s also in the case of $2+1$
  dimensions \cite{Svetitsky,SvetitskyReview}.

One may wish, as we do here for simulation purposes, to consider the case of a finite, truncated, local Hilbert space. In this case, the strong limit behavior is unaffected. For the weak limit, only when it is confining one can
approximate the behavior of the real theory - as in the confining phase only a few number of low flux states is needed. The approximation becomes better as the truncation value $\ell$ is stronger, but also for small values of
$\ell$ a reasonable approximation holds \cite{Zohar2012}. This is true when one considers the lattice theory; In terms of the continuum limit, if the truncation is done properly one can regain the right continuum limit - see,
for example, the link models \cite{Horn1981,Orland1990,Wiese1997,Brower1999}.

\subsection*{2. Dynamic fermions in cQED}

Including dynamic fermions in a lattice gauge theory is a bit problematic, due to the doubling of fermions in the continuum limit. We focus, in this paper, and thus here, in the method of staggered fermions \cite{Susskind1977}.
There, the continuum spinors decompose into two neighboring vertices of the lattice, one containing particles and the other - antiparticles. 2-component continuum spinors (which are sufficient for $1+1$ dimensional theories,
and also for $2+1$ dimensional ones, if one is not intersted in the chiral anomaly) are thus formed of two neighboring vertices, containing up to a single fermion each. This is why the $H_M$ part of the Hamiltonian (\ref{MQED})
contains the masses with alternating signs: the odd vertices represent the antiparticles, and they are filled in the ground state - forming the Dirac sea. Moreover, the dynamic charge $Q_{\mathbf{n}}$, present in Gauss's law
(\ref{gausslaw}) is defined as
\begin{equation}
Q_{\mathbf{n}} = \psi_{\mathbf{n}}^{\dagger}\psi_{\mathbf{n}} - \frac{1}{2}\left(1-\left(-1\right)^{n_1+n_2}\right)
\end{equation}
(for simplicity this is a $2+1$ dimensional definition, which can be generalized to other dimensions, keeping the vertices' parity properties). This is in with the particle/anti-particle picture: even vertices, representing
 particles, can be occupied by no fermions - corresponding to no mass and no charge - or a single fermion, corresponding to a particle with mass $M$ and charge $+1$; Occupied odd vertices have no mass (relative to $-M$) and no
 charge,
  and correspond to vacant vertices in the HEP picture, while empty vertices have mass $M$ (relative to $-M$) and charge $-1$, representing an anti-particle.

The fermionic interactions are obtained with the gauge invariant $H_{int}$, whose abelian form is (\ref{abhint}). However, one should note that in order to get the Dirac equation in the continuum limit, some phases should
 introduced to this Hamiltonian, and they can be achieved using canonical transformations (which are not required to be implemented experimentally) on the fermions. For example, in a $1+1$ dimensional system, one can use the
 transformation in \footnotemark[5]. The phase prescription for $3+1$ dimensions, yielding four-component spinors, is given in \cite{Susskind1977}.

\subsection*{3. Confinement in $\mathbb{Z}_N$ gauge theories}

Confinement for these theories is especially interesting, as it was argued \cite{Hooft1978} that confinement of quarks in $QCD$ ($SU(3)$) has to do with the group's center, which is the group $\mathbb{Z}_3$.
Thus the phase structure of $\mathbb{Z}_N$ is interesting \cite{Elitzur1979,Horn1979,Bhanot1980}.
The theory confines in the strong coupling limit ($\mu \rightarrow \infty$). Both in $2+1$ and in $3+1$ dimensions it in not the only phase. In $3+1$ dimensions, due to the self-dual nature of the theory, there is a phase transition
from electric confinement for large coupling to a magnetic confinement in the small coupling regime, for $N < N_C$ ($N_c \approx 6$). For $N > N_C$ there is a third phase, with no confinement at all.
In $2+1$ dimensions there is a phase transition and the theory does not confine in the weak limit \cite{Horn1979}. However, for $N \rightarrow \infty$ the theory shows the phase transition at $g=0$ \cite{Bhanot1980}, in accordance
with the
single confining phase structure of $U(1)$ (the $N \rightarrow \infty$ limit of $\mathbb{Z}_N$).

\subsection*{4. $SU(N)$ gauge theories - dynamic fermions and confinement}

In order to introduce dynamical fermions, we choose to utilize the method of staggered fermions again.
We choose, as usual, to suppress flavor and spin indices, and identify the spinor components only by group indices, as in equation (\ref{intspin}). With staggered fermions the masses are positive/negative alternately, similarly to (\ref{MQED}). This time, the non-abelian charge is defined as
\begin{equation}
\left(Q_{\mathbf{n}}\right)_a =
\underset{bc}{\sum} \left(\psi_{\mathbf{n}}^{\dagger}\right)_b \left(T_a\right)_{bc}\left(\psi_{\mathbf{n}}\right)_c =
\frac{1}{2}\underset{bc}{\sum} \left(\psi_{\mathbf{n}}^{\dagger}\right)_b \left(\sigma_a\right)_{bc}\left(\psi_{\mathbf{n}}\right)_c
\end{equation}
(where the second equality holds for the fundamental representation).
This is a fermionic Schwinger representation, and thus a full vertex has a zero charge. Thus one can establish a Dirac sea, similarly the abelian one (appendix 1), where vacancies in odd vertices correspond to anti-particles, measuring the masses there relative to $-M$ per fermion.

One can also consider static charges (i.e. with infinite mass). In
particular, one would like to examine the confinement of static fundamental
charges. This was initially done by Wilson \cite{Wilson} and later
other authors in the Euclidean approach \cite{Polyakov,BanksMyersonKogut},
but we shall concentrate on the Hamiltonian picture. Consider
the strong limit of the theory, $g^{2}\gg1$, in which the electric
part is much stronger than the
magnetic part, which may be thus considered as a perturbation (as in the abelian case). In
this context, the ground state of two charges in $\mathbf{n},\mathbf{n}+R\mathbf{\hat{x}}$,
is the non-abelian analogue to (\ref{fluxtube}):
Thus, the links along the straight line connecting the charges are
excited to $j=\frac{1}{2}$, contributing zeroth order energy $\frac{g^{2}}{2}j\left(j+1\right)=\frac{3}{4}\frac{g^{2}}{2}$,
while the other links are in the singlet $j=0$, contributing no energy
at all in zeroth order. Therefore the energy of this state, to zeroth order,
is $E\left(\phi\right)=\frac{3}{4}\frac{g^{2}}{2}R$ (as there are
$R$ excited links), and we get confining behaviour, $V\left(R\right) \propto R$ again: the static energy (potential)
between two static {}``quarks'' is proportional to their distance.
The string tension, $E/R$, is $\frac{3}{4}\frac{g^{2}}{2}$ in this
case \cite{Kogut1983}. Other coupling
regimes will not be treated here. We shall only comment that
as the coupling decreases within a confining phase, the flux tube
broadens and $\left|\phi\right\rangle $ is no longer an eigenstate
of the Hamiltonian, in accordance with the growth of perturbative
corrections, until the breakdown of perturbation theory.

\bibliography{ref}

\end{document}